\documentclass[twocolumn]{aastex63}
\usepackage{amsmath, amsthm, amssymb, amsfonts}
\usepackage[T1]{fontenc}
\usepackage{natbib}
\usepackage{graphicx}
\usepackage{float}
\usepackage{cleveref}
\usepackage{xcolor}
\defcitealias{poggio}{P20}
\defcitealias{martin}{LS19}
\defcitealias{cheng}{C20}
\usepackage[utf8]{inputenc}
\usepackage{rotating}

\usepackage{savesym} 
\savesymbol{tablenum} 
\usepackage{siunitx}

\restoresymbol{SIX}{tablenum}

\begin{document}
	
	\title{A Case against a Significant Detection of Precession in the Galactic Warp}
	
	\author{\v{Z}. Chrob\'{a}kov\'{a}}
	\affiliation{Instituto de Astrofísica de Canarias, E-38205 La Laguna, Tenerife, Spain}
	\affiliation{Departamento de Astrofísica, Universidad de La Laguna, E-38206 La Laguna, Tenerife, Spain}
	
	\author{M. L\'{o}pez-Corredoira}
	\affiliation{Instituto de Astrofísica de Canarias, E-38205 La Laguna, Tenerife, Spain}
	\affiliation{Departamento de Astrofísica, Universidad de La Laguna, E-38206 La Laguna, Tenerife, Spain}
	
	\begin{abstract}
		Recent studies of warp kinematics using Gaia DR2 data have produced detections of warp precession for the first time, which greatly exceeds theoretical predictions of models. However, this detection assumes a warp model derived for a young population (few tens of megayears) to fit velocities of an average older stellar population of the thin disk (several gigayears) in Gaia-DR2 observations, which may lead to unaccounted systematic errors.
		Here, we recalculate the warp precession with the same approach and Gaia DR2 kinematic data, but using different warp parameters based on the fit of star counts of the Gaia DR2 sample, which has a much lower warp amplitude than the young population. When we take into account this variation of the warp amplitude with the age of the population, we find that there is no need for precession. We find the value of warp precession $\beta = 4^{+6}_{-4}$ km s$^{-1}$ kpc$^{-1}$, which does not exclude nonprecessing warp.
	\end{abstract}
	
	
\section{Introduction}\label{intro}
Although the warp was discovered over 60 yr ago \citep{kerr,oort} and extensively studied in the past decades \citep[and others]{carney,martin_warp,amores,chen}, the mechanism causing the warp is still unknown. Theories include accretion of intergalactic matter onto the disk \citep{martin_accretion}, interaction with other satellites \citep{kim}, the intergalactic magnetic field \citep{battaner}, a misaligned rotating halo \citep{debattista}, and others. 

Kinematic studies are necessary in order to reveal the origin of the warp, as different models give different predictions. Warp precession is estimated by various mechanisms of warp formation, such as torque of the dark halo on the disk \citep{dubinski}, torque on the disk due to misalignment of the halo and the disk \citep{jiang}, or accretion of intergalactic matter onto the disk \citep{martin_accretion,jeon}.

Recently, studies of warp kinematics and time evolution were made in order to understand the origin of the warp \citep{poggio_kin,poggio,haifeng}. In particular, the result of \citet[hereafter P20]{poggio} is of great interest as they measured the precession of the Galactic warp for the first time. Using the second Gaia data release DR2 \citep{gaia1,gaia} combined with Two Micron All-Sky Survey \citep[2MASS]{2mass} photometry, they find the warp precession to be $\beta=10.86 \pm 0.03 (stat.) \pm 3.20 (syst.)$ km s$^{-1}$ kpc$^{-1}$. Their result was supported by another recent work of \cite{cheng}, who measured a similar value of precession $\beta=13.57^{+0.20}_{-0.18}$ km s$^{-1}$ kpc$^{-1}$ by fitting vertical velocities using data from Gaia DR2 and Apache Point Observatory Galactic Evolution Experiment 2 \citep[APOGEE-2]{majewski}, as contained in Sloan Digital Sky Survey (SDSS) Data Release 16 \citep[DR16]{apogee}. These results greatly exceed predictions of warp precession estimated considering various mechanisms of warp formation, such as torque of the dark halo on the disk \citep{dubinski}, torque on the disk due to misalignment of the halo and the disk \citep{jiang}, or accretion of intergalactic matter onto the disk \citep{martin_accretion,jeon}. All these works predict a small value of precession, between $0.1$ and $1.5$ km s$^{-1}$ kpc$^{-1}$. The result of \citetalias{poggio} contradicts conclusions of dynamical models of warp formation and suggests that warp is a transient response of the disk to an outside interaction rather than a slowly evolving structure. We know that most spiral galaxies have warps \citep{reshetnikov,sanchez}, so it seems unlikely that the warp is caused by a transient phenomenon in all of them unless they all have a substructure that triggers warp formation very often.

Kinematics studies also discovered significant substructures in the disk velocities \citep{haifeng_maps,martin_new,xu}. In \cite{martin_new} it is shown that the features in vertical motions can be partially explained by the warp; however, the main observed features can only be explained in terms of out-of-equilibrium models. Nevertheless, our purpose in this paper is not to explain all the features of the observed velocities, but just to verify that a simple warp model without precession is able to fit the data, against the claims of \citetalias{poggio}. Therefore, we will consider the substructures as random fluctuations and focus only on the large structural effect of the warp.

In this work, we challenge the result of \citetalias{poggio}, as we think they did not consider the variation of the amplitude of the warp with the age of the population, and present our own calculation of the warp precession rate. \citetalias{poggio} derived the warp precession under an assumption that the amplitude of the warp corresponding to specific stellar populations (Classical Cepheids, Pulsars) is representative of the whole population. This is not consistent with data, as the stellar population traced by the Gaia survey is significantly older on average ($\sim$5-6 Gyr) and it presents a much lower amplitude of the warp than the young population one \citep{zofi,haifeng}. For simplicity, we will refer to this population as an ``old population'' and to Cepheids as a ``young population.'' We show that when applying the warp model for an old population we cannot detect the precession of the warp; therefore, we cannot favor any model of warp formation.
	
\section{Data selection}
We use the data of \citet[hereafter LS19]{martin}, who have produced extended kinematic maps of the Milky Way by using Gaia DR2, by considering stars with measured radial heliocentric velocities and with parallax errors less than $100\%$; the total sample contains 7,103,123 sources. Such objects were observed by the Radial Velocity Spectrometer \citep[RVS;][]{cropper} that collects medium-resolution spectra (spectral resolution $\frac{\lambda}{\Delta \lambda}\approx 11 700$) over the wavelength range 845-872 nm, centered on the Calcium triplet region. Radial velocities are averaged over a 22 month time span of observations. Most sources have a magnitude brighter than 13 in the $G$ filter.

In more detail, the effective temperatures for the sources with radial velocities that \citetalias{martin} have considered are in the range of 3550 to 6900 K. The uncertainties of the radial velocities are 0.3 km s$^{-1}$ at $G_{\mathrm{RVS}} < 8$, 0.6 km s$^{-1}$ at $G_{\mathrm{RVS}}=10$, and 1.8 km s$^{-1}$ at $G_{\mathrm{RVS}}= 11.75$; plus systematic radial velocity errors of $< 0.1$ km s$^{-1}$ at $G_{\mathrm{RVS}} < 9$ and 0.5 km s$^{-1}$ at $G_{\mathrm{RVS}}= 11.75$. The uncertainties of the parallax are 0.02 –0.04 mas at $G<15$, 0.1 mas at $G=17$, 0.7 mas at $G=20,$ and 2 mas at $G=21$. The uncertainties of the proper motion are 0.07 mas $\mathrm{yr}^{-1}$ at $G<15$, 0.2 mas $\mathrm{yr}^{-1}$ at $G=17$, 1.2 mas $\mathrm{yr}^{-1}$ at $G=20,$ and 3 mas $\mathrm{yr}^{-1}$ at $G=21$. For details on radial velocity data processing and the properties and validation of the resulting radial velocity catalog, see \cite{sartoretti} and \cite{katz}. The set of standard stars that was used to define the zero-point of the RVS radial velocities is described in \cite{soubiran}. \citetalias{martin} consider the zero-point bias in the parallaxes of Gaia DR2, as found by \cite{lindegren}, \cite{arenou}, \cite{stassun}, and \cite{zinn}; however, they find that the effect of the systematic error in the parallaxes is negligible. 

As the parallax error grows as the distance from us grows, \citetalias{martin} applied a statistical deconvolution of the parallax errors based on Lucy's inversion method \citep{lucy} to statistically estimate the distance. In this way they have derived the extended kinematical maps in the range of Galactocentric distances up to 20 kpc.

From this sample, we only choose stars with the Galactic latitude $\lvert b \rvert<\ang{10}$, Galactocentric distance $R>12$ kpc, and heliocentric distance $r<8$ kpc, so we obtain a dataset with Galactocentric radius $12<R<16$ kpc, maximum vertical distance $z\approx1.4$ kpc, and within $\lvert \phi \rvert<\ang{40},$  as seen from the Galactic center. We bin the data in Galactocentric Cartesian coordinates with size $\Delta X=1.0$ kpc and $\Delta Y=1.0$ kpc. For $R<12$ kpc, the amplitude of the warp is too small to be considered and the relative error of the warp model is higher. For $r>8$ kpc, the errors of vertical velocities are very large.

\section{Methods}\label{methods}
We model the warp in Galactocentric cylindrical coordinates $(R,\phi,z)$ using the warp model of \citet[Equation (11)]{zofi} for which the average height over the plane of the Galactic disk stars $z_w$ is calculated as

\begin{eqnarray}\label{9}
z_w=[C_wR(pc)^{\epsilon_w}sin(\phi-\phi_w)+17]~pc~.
\end{eqnarray}
The model describes the warp as a series of tilted rings, where the amount of tilt is the Galactocentric distance raised to a power of $\epsilon_w$, $C_w$ is the warp amplitude, and $\phi_w$ is the Galactocentric angle defining the warp’s line of nodes. $C_w,\epsilon_w$, and $\phi_w$ are free parameters of the model, which were fitted in \cite{zofi}. The 17 pc term compensates for the elevation of the Sun above the plane \citep{z_slnko}. We consider time evolution of the warp amplitude and warp precession

\begin{eqnarray} 
C_w(t)&=&C_{w,max}sin(\omega t+ \alpha)~, \label{param1} \\
\phi_w(t)&=&\phi_{w0}+\beta t~. \label{param2}
\end{eqnarray}

The values of model parameters given by \cite{zofi} are
\begin{eqnarray}\label{parametre}
C_w&=&1.17\cdot10^{-8} \mathrm{pc} \pm 1.34\cdot10^{-9} \mathrm{pc} (stat.) \nonumber \\
&\phantom{}&^{+0 \mathrm{pc}}_{-2.9\cdot10^{-10} \mathrm{pc}}(syst.)~, \nonumber \\
\epsilon_w&=&2.42\pm 0.76(stat.) ^{+0.129}_{-0} (syst.)~, \\
\phi_w&=&\ang{-9.31}\pm \ang{7.37} (stat.) ^{+\ang{4.48}}_{-\ang{0}} (syst.)~. \nonumber
\end{eqnarray}

We follow the approach of \citetalias{poggio}, taking the zeroth moment of the collisionless Boltzmann equation to derive the expression for vertical velocity.

\begin{eqnarray}\label{model}
v_z(R,\phi)&=&C_{w,0}K R^{\epsilon_w}sin(\phi-\phi_{w,0}) \nonumber \\
&-&C_{w,0}R^{\epsilon_w}\beta cos(\phi-\phi_{w,0}) \\
&+&C_{w,0}R^{\epsilon_w-1}cos(\phi-\phi_{w,0})v_{\phi}~, \nonumber
\end{eqnarray}	
where $C_{w,0}$ and $\phi_{w,0}$ are Eqs. (\ref{param1}) and (\ref{param2}) at present time $t$ and

\begin{eqnarray}
K=\omega \cdot cotan(\alpha)~.
\end{eqnarray}
We use the least-squares method, to fit the data, minimizing $\chi^2$: 
\begin{eqnarray}
{\chi^2=\sum_{i} \frac{(v^{obs}_{Z,i}-v^{model}_{Z,i})^2}{\sigma^2_i}~.}
\end{eqnarray}
\cite{martin} give their data in coordinates $(X,Y)$ binned with size $\Delta X=0.2$ kpc and $\Delta Y=0.2$ kpc. To avoid confusion, we will refer to those bins as pixels. Each bin with average velocity $v^{obs}_{Z,i}$ is generated by grouping different pixels $(X,Y)$ into squares of $1$ kpc $\times 1$ kpc with the same number of pixels for all bins, where $\sigma^2_i$ is the error of velocity in each bin:

\begin{eqnarray}\label{err}
{\sigma^2_i=\frac{\sum_{j\in i} w_j \sigma_{Z,j}^2}{\sum_{j \in i} w_j}~,}
\end{eqnarray}
where $\sigma_{Z,j}$ is the error of velocity in the pixel $j$ and $w_j$ is its weight (number of stars in our case).

Errors in the bin can be calculated differently, depending on whether the statistical or systematic error dominates. We tried to calculate the error using different approaches, but we find that the results are similar in all cases. More about error estimation can be found in Section \ref{errors}.

\section{Results}
\subsection{Old Population Warp Does Not Give Precession}
\begin{figure}
	\begin{center}
		
		\includegraphics[width=0.5\textwidth]{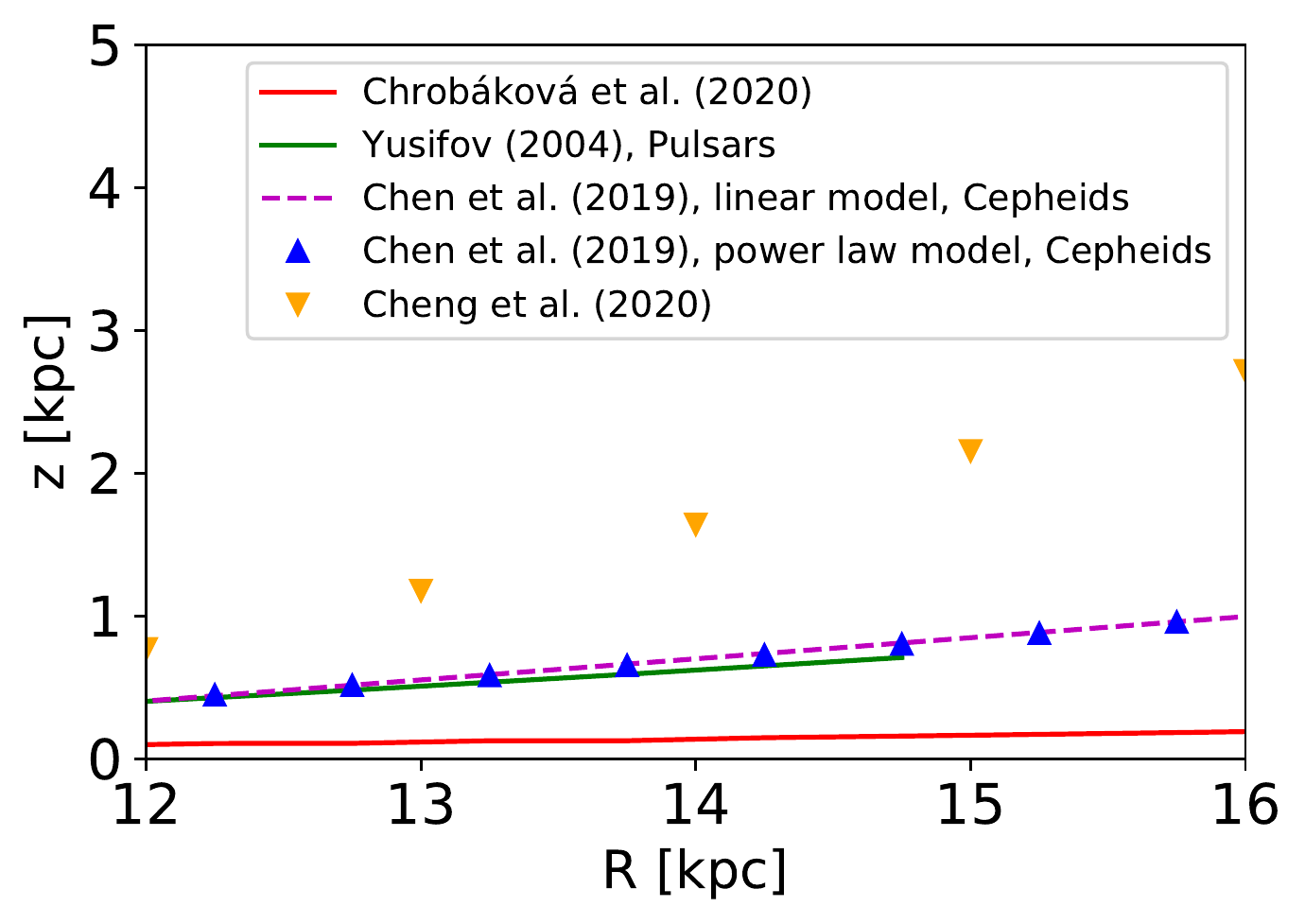}
		\caption{Comparison of maximum amplitudes of various warp models, used to calculate the precession. It is evident that the model of \cite{zofi}, derived for the whole stellar population, reaches a significantly lower amplitude than other works.}\label{modely_porovnanie}
	\end{center}
\end{figure}

\begin{figure*}
		\gridline{\fig{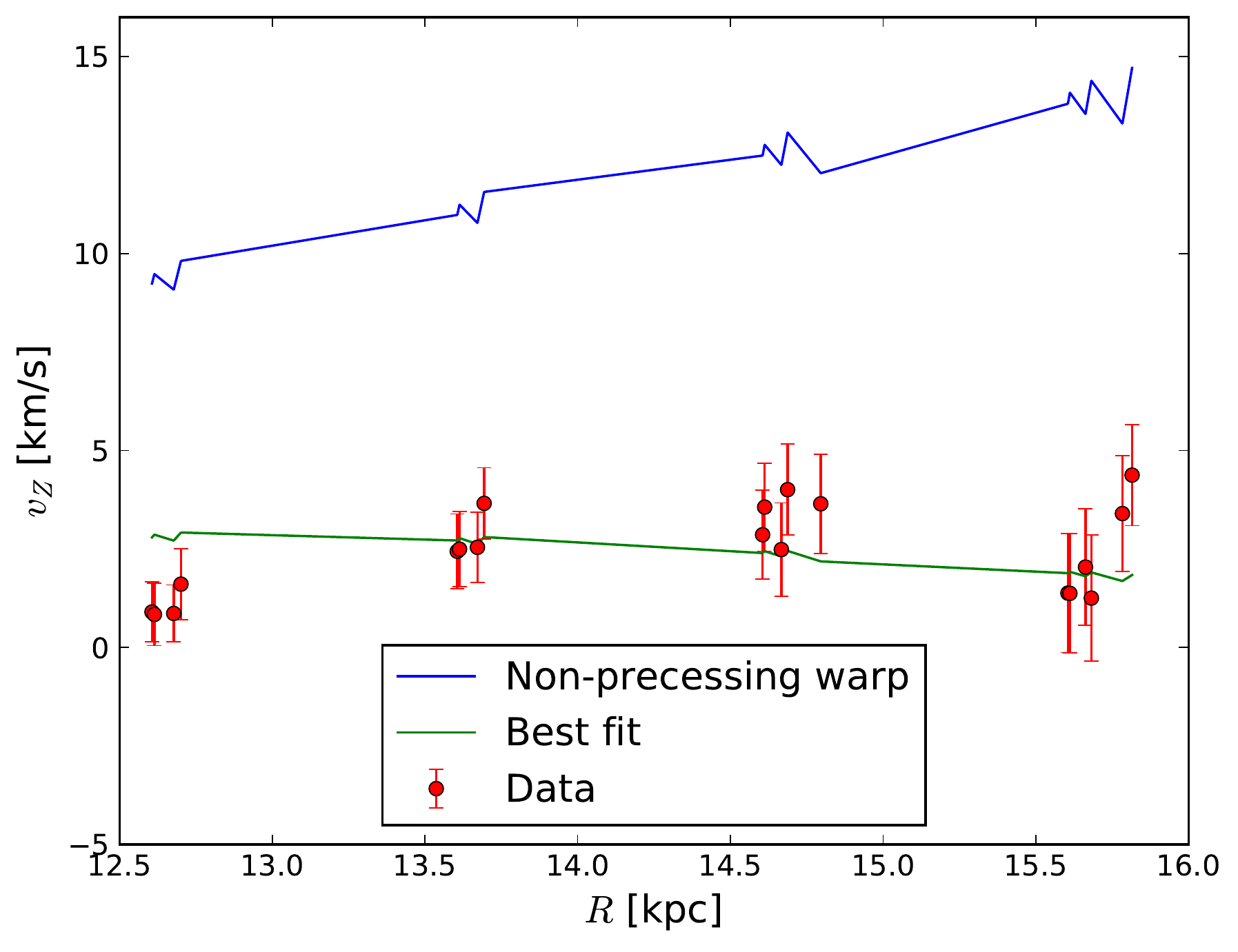}{0.4\textwidth}{(a)}
		\fig{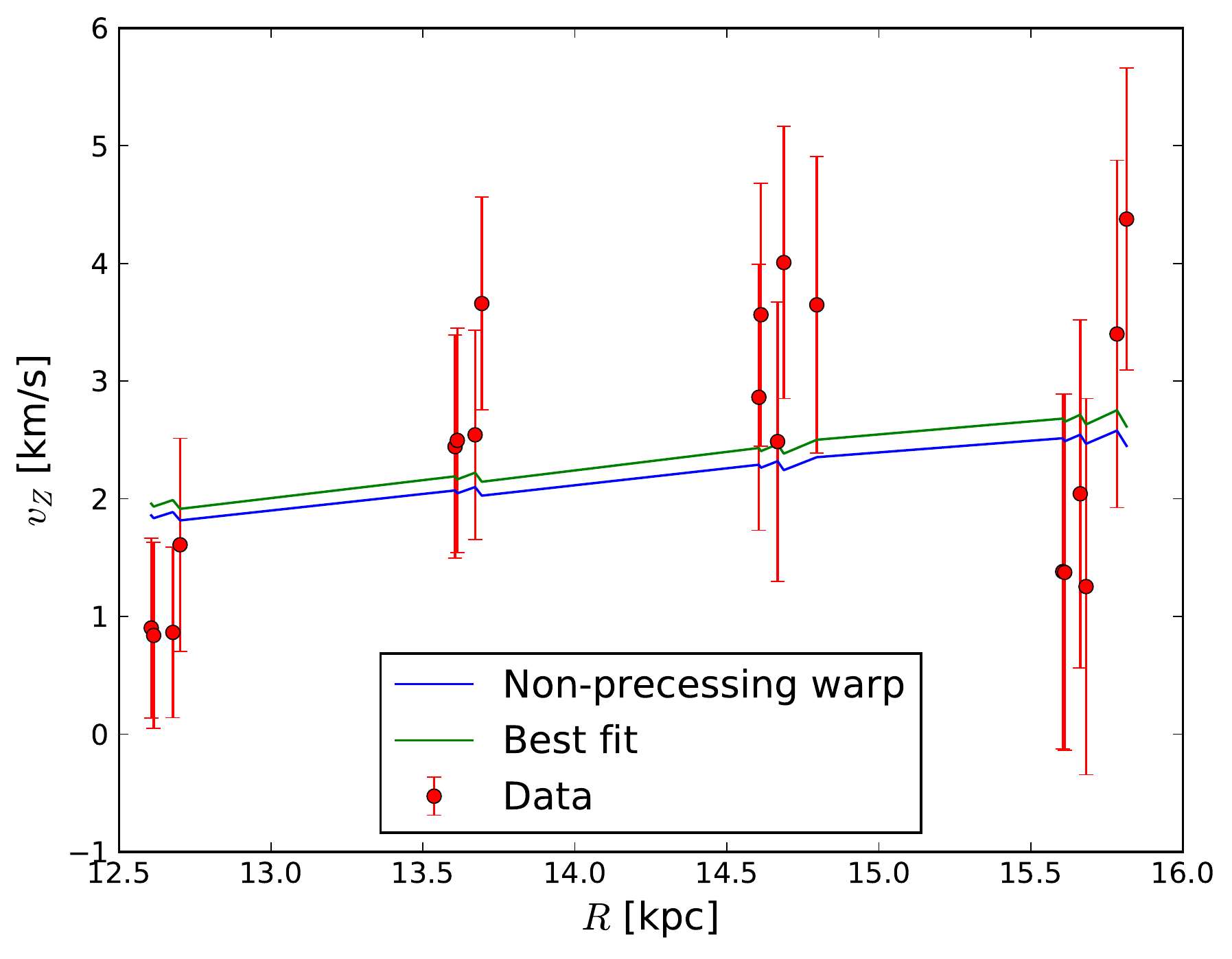}{0.4\textwidth}{(b)}
	}
		\gridline{
		\fig{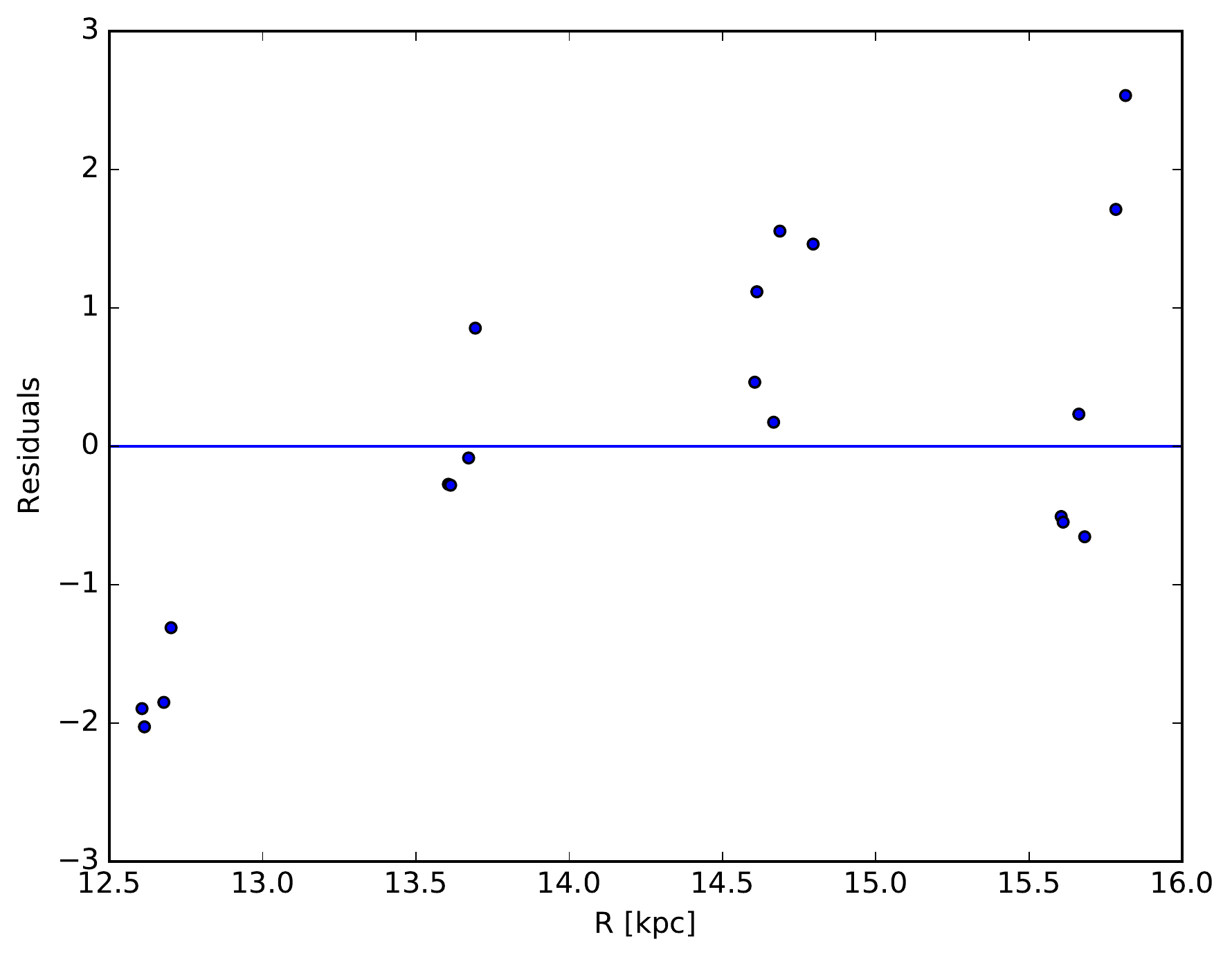}{0.4\textwidth}{(c)}
		\fig{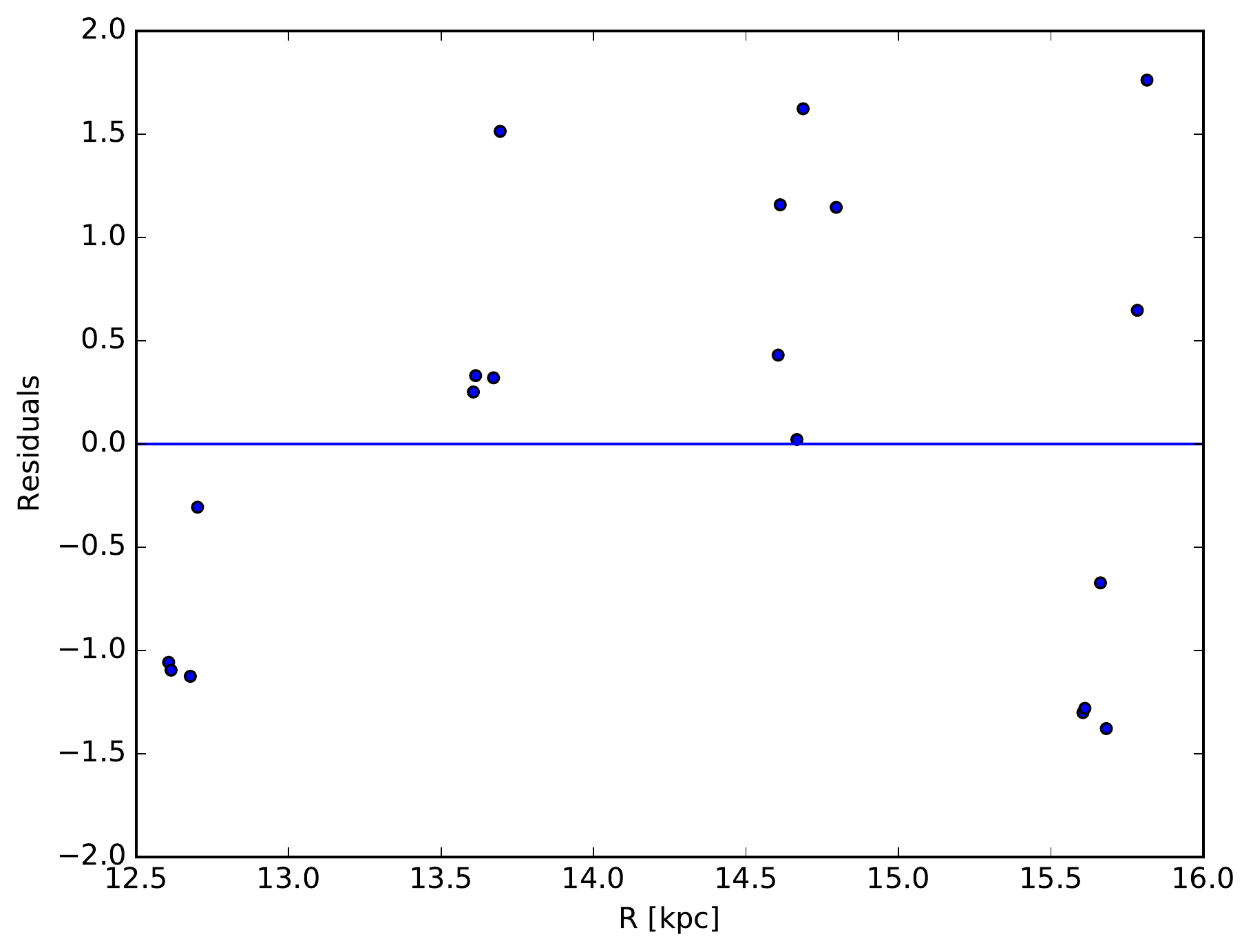}{0.4\textwidth}{(d)}
	}

	\caption{Result of the fit with warp models, using only data near the anticenter. The blue curve is for the nonprecessing model of the warp ($\beta=0$ km s$^{-1}$ kpc$^{-1}$), the green curve is the best fit of the data. (a) Model of \citetalias{poggio}, with the warp parameters of \cite{chen}. For the best fit, we find $\beta=13 \pm 1$ km s$^{-1}$ kpc$^{-1}$. (b) Model of \cite{zofi}. For the best fit, we find $\beta=-1 \pm 9$ km s$^{-1}$ kpc$^{-1}$. (c) Residuals corresponding to the model of \citetalias{poggio}, with warp parameters of \cite{chen}. (d) Residuals corresponding to the model of \cite{zofi}.}\label{lon}
\end{figure*}

We fit the vertical velocities with a warp model considering both the evolution in time of the warp amplitude and the warp precession. As a first case, we only take data with azimuth $\lvert \phi \rvert<\ang{10}$ (near the anticenter), where the dependence of amplitude on time is negligible, as can be seen from Equation (\ref{model}), when we set the value of the angle of the line of nodes. Therefore we only have one free parameter $\beta$ that represents precession of the warp azimuth of the line of nodes (see Equation (\ref{param2})). We carry out the fit for the model of \cite{zofi} and the model of \citetalias{poggio} using the warp parameters of the linear model of \cite{chen}. Differences in maximum amplitude for these warp models can be seen in Figure \ref{modely_porovnanie}. In Figure \ref{lon}(a) we show the best fit and the nonprecessing warp for the model of \citetalias{poggio}. It is clear that, with the high amplitude, the nonprecessing model gives much higher velocity than what is observed, so a high precession is necessary to compensate for that. That is precisely the result of \citetalias{poggio}. In contrast, the warp model representative of the old stellar population of Gaia DR2 \citep{zofi} reaches about 3\textendash4 times lower amplitude than the one used by \citetalias{poggio} (see Figure \ref{modely_porovnanie}); therefore, it does not need a high precession.
 
As shown in Figure \ref{lon}(b), with a nonprecessing warp the amplitude of the velocity is already of the order of the data and the precession gives only a small correction. The value of precession we obtain with the model used by \citetalias{poggio} is $\beta=13 \pm 1$ km s$^{-1}$ kpc$^{-1}$, which is similar, although a bit higher than the value obtained by \citetalias{poggio} ($\beta=10.86 \pm 0.03 (stat.) \pm 3.20 (syst.)$ km s$^{-1}$ kpc$^{-1}$), probably due to difference in datasets. The value of precession using the old stellar population of Gaia DR2 is $\beta=-1 \pm 9$ km s$^{-1}$ kpc$^{-1}$, which is consistent with the nonprecessing warp model, as well as with the result of \citetalias{poggio}, since we have large error bars. \citetalias{poggio} already pointed out that a warp model with a small amplitude like ours would fit the data without the need for precession; however, they do not carry out the analysis with such a model and do not provide a result of such fit; therefore, we cannot compare our result with theirs directly.

\begin{figure*}
	\gridline{\fig{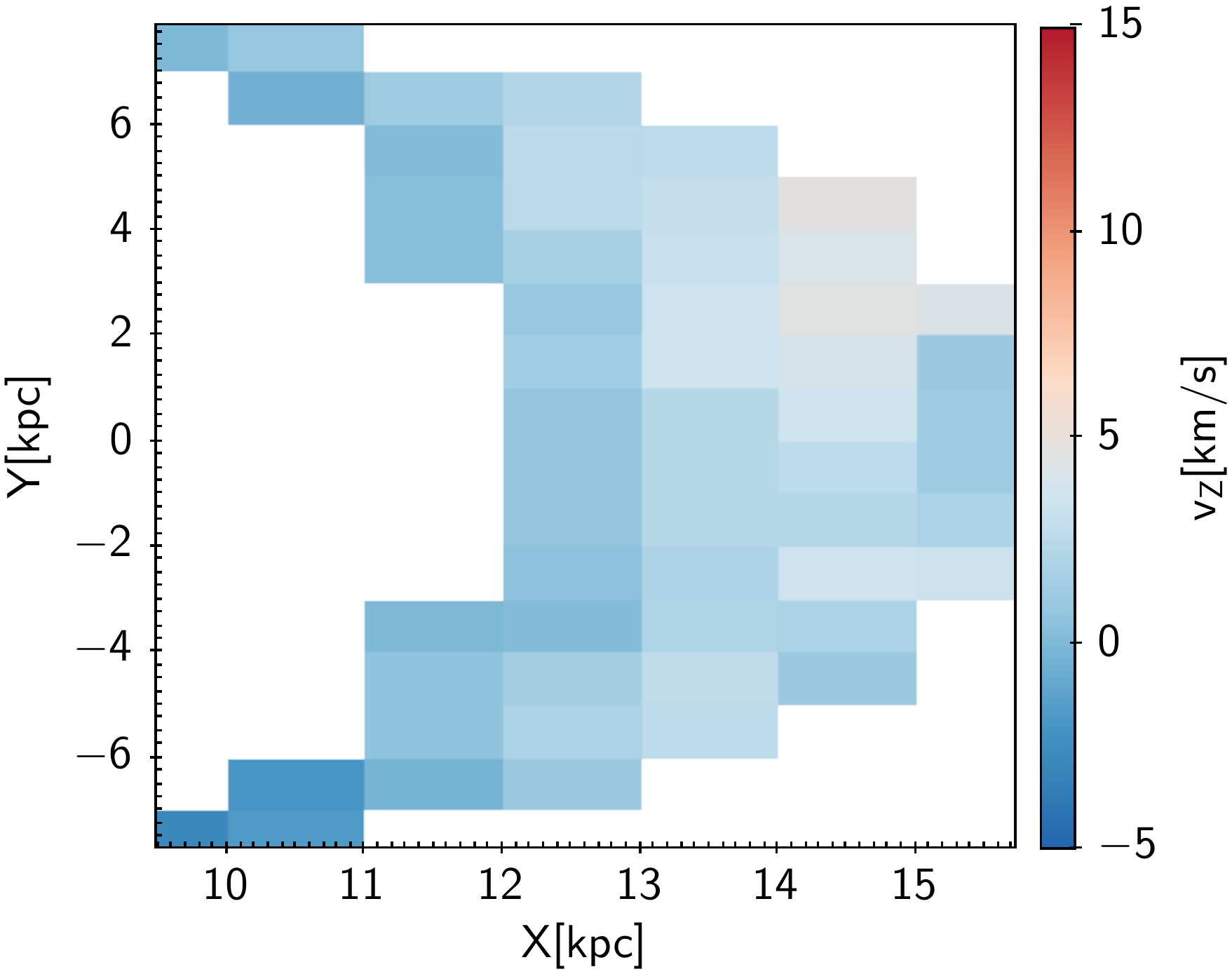}{0.35\textwidth}{(a)}
		\fig{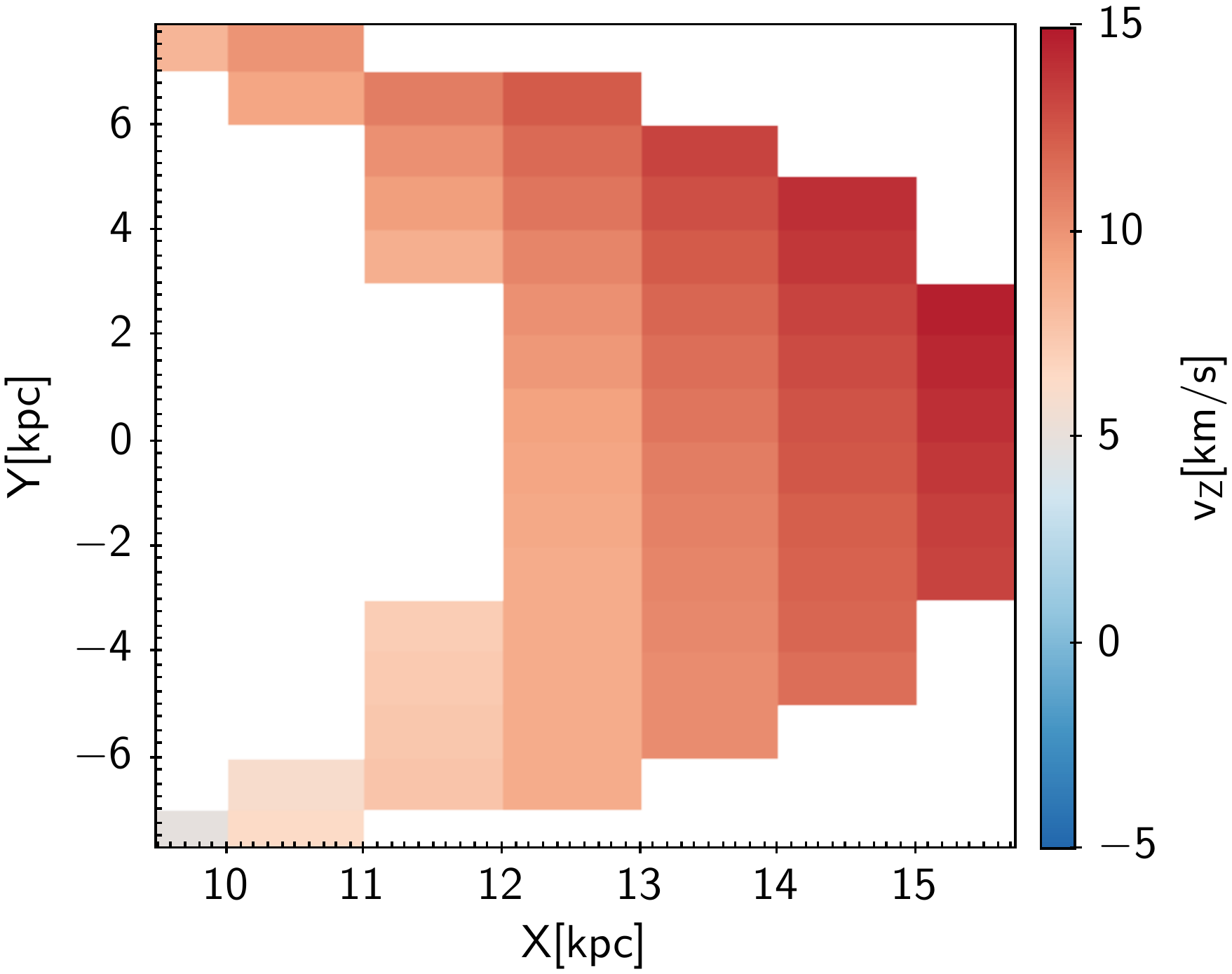}{0.35\textwidth}{(b)}
		\fig{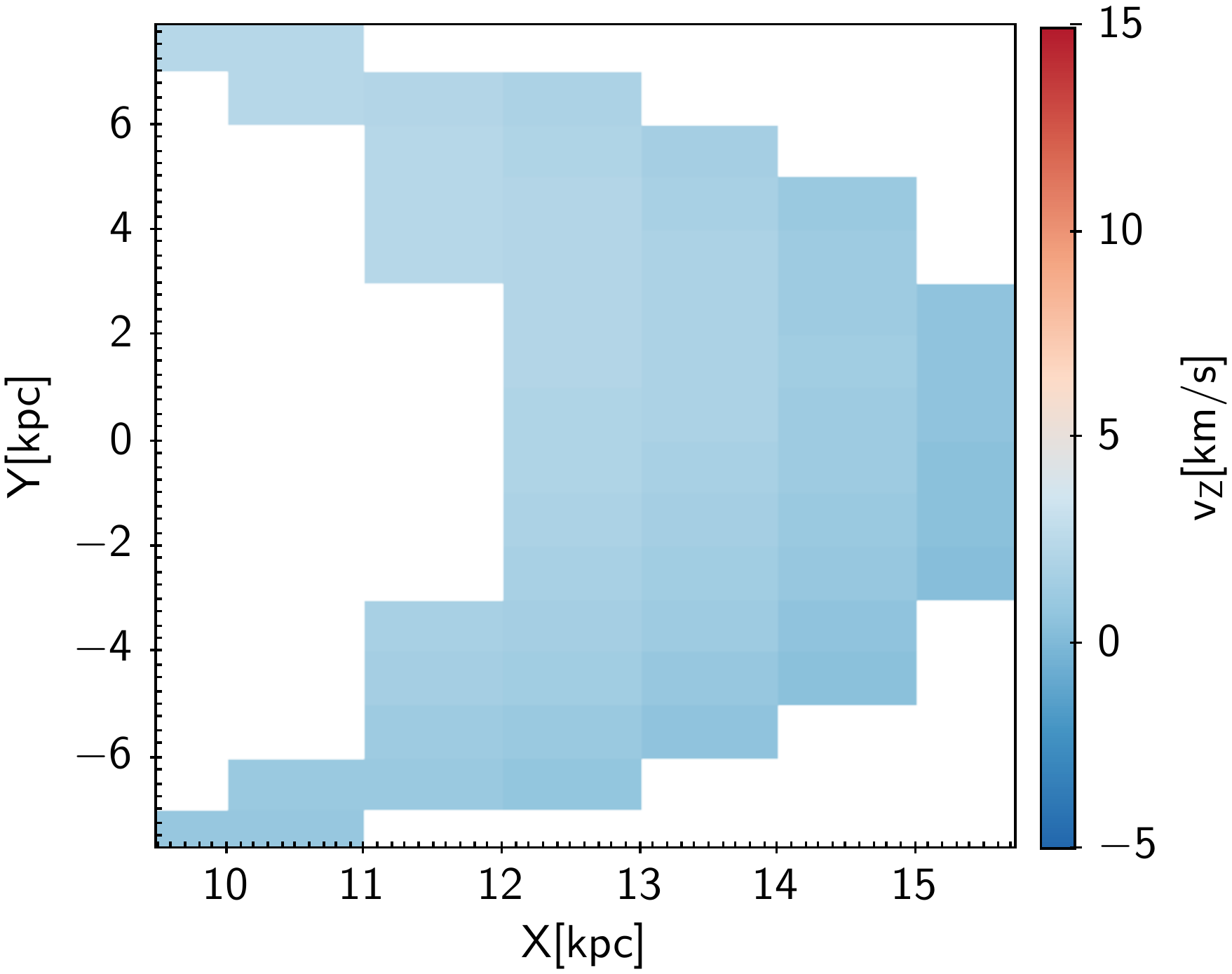}{0.35\textwidth}{(c)}
	}
		
	\gridline{\fig{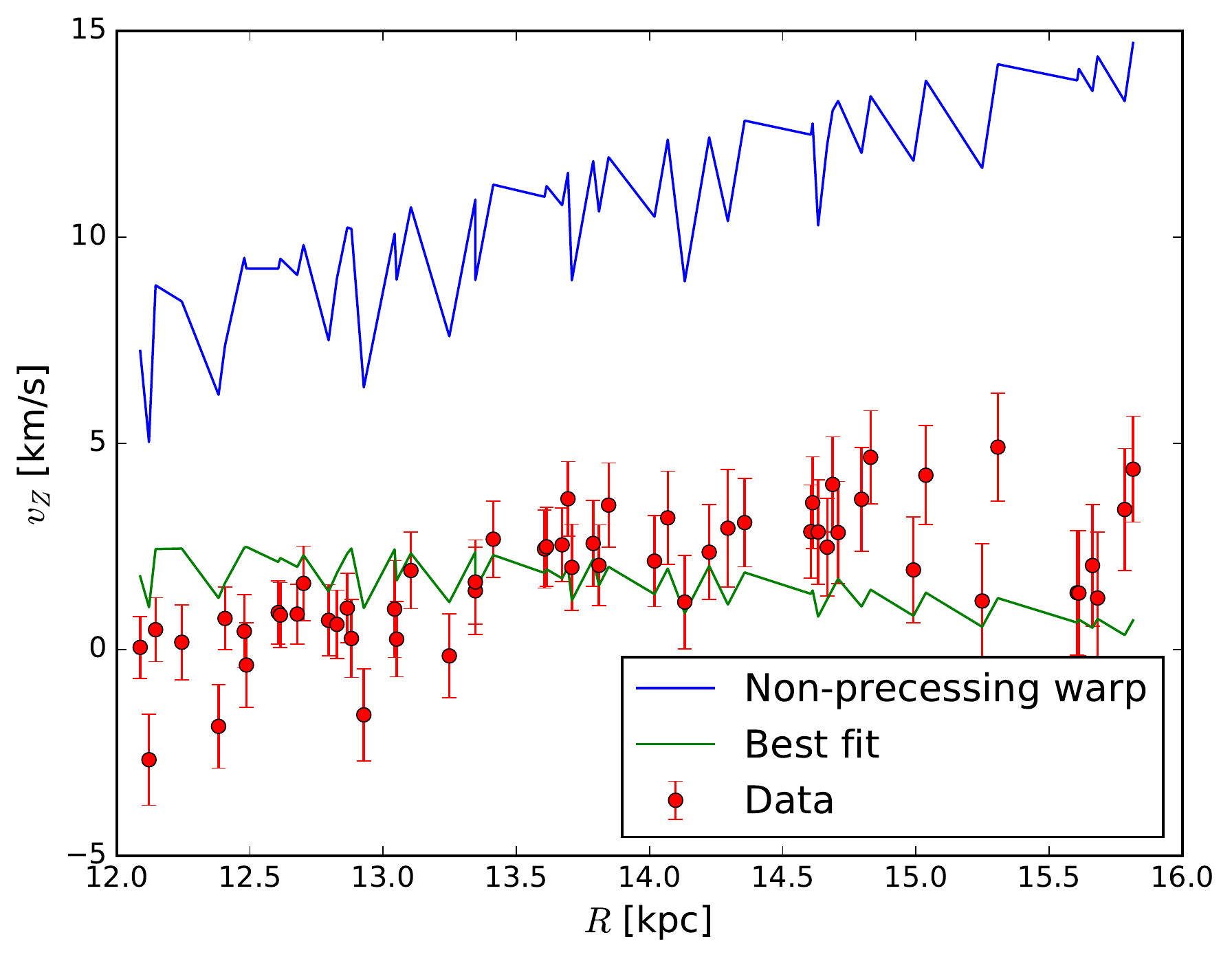}{0.35\textwidth}{(d)}
	\hspace{-2cm}
	\fig{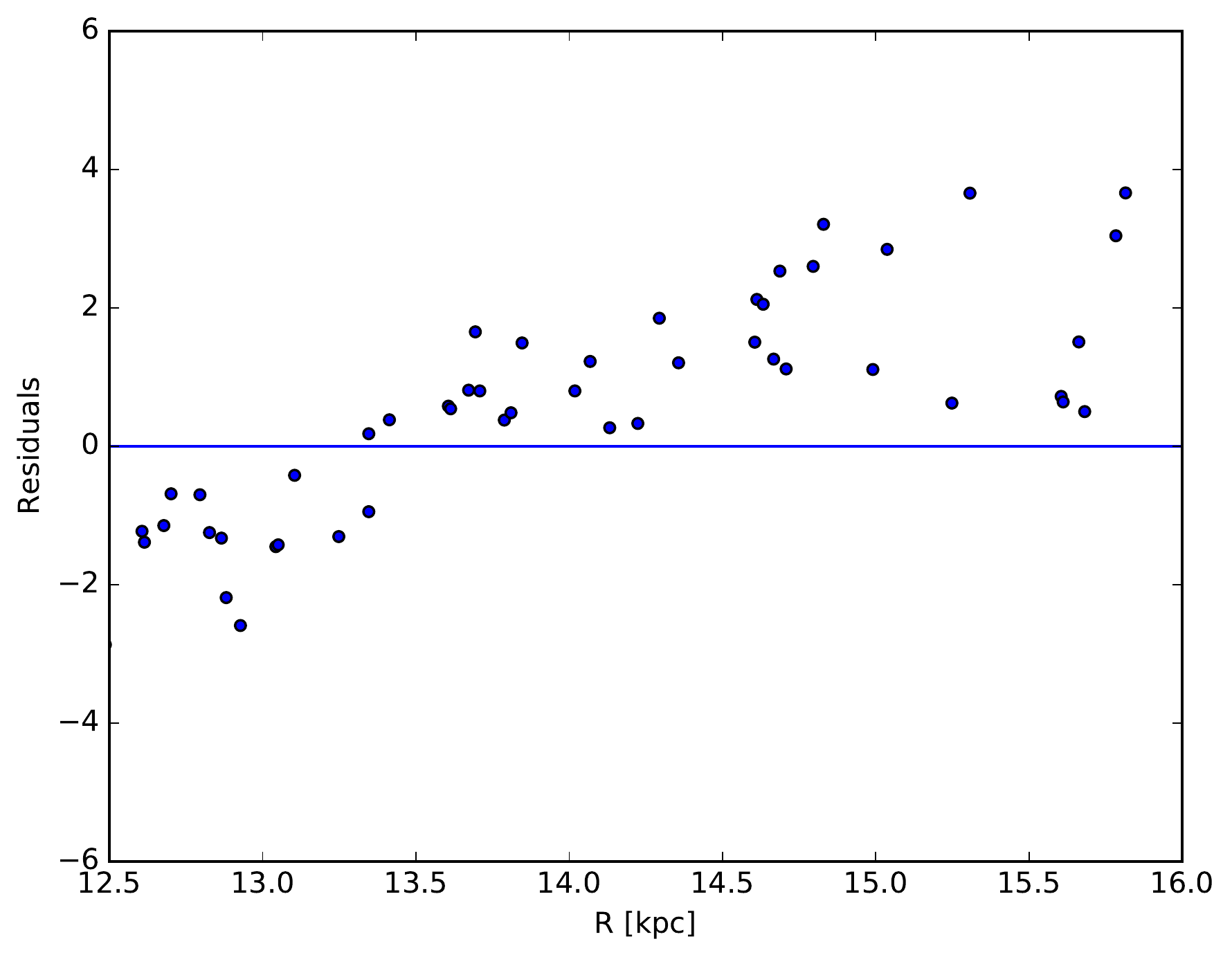}{0.35\textwidth}{(e)}
	}
	
	\caption{Result of fit using Equation (\ref{model}) with parameters of \cite{chen}, fitting all data. (a) Data. (b) Nonprecessing steady warp ($\beta=0$ km s$^{-1}$ kpc$^{-1}$ and $K=0$ km s$^{-1}$ kpc$^{-1}$). (c) Best fit with $K$ and $\beta$ as free parameters. For the best fit, we find $K=1.0^{+1.4}_{-2.2}$ km s$^{-1}$ kpc$^{-1}$ and $\beta=14.0 \pm 1.4$ km s$^{-1}$ kpc$^{-1}$. (d) Two-dimensional representation of the fit. The blue curve is for the nonprecessing steady model of the warp, the green curve is the best fit of the data. (e) Residuals corresponding to the best fit.}
		\label{fig_final_chen}
\end{figure*}

\begin{figure*}
	\gridline{\fig{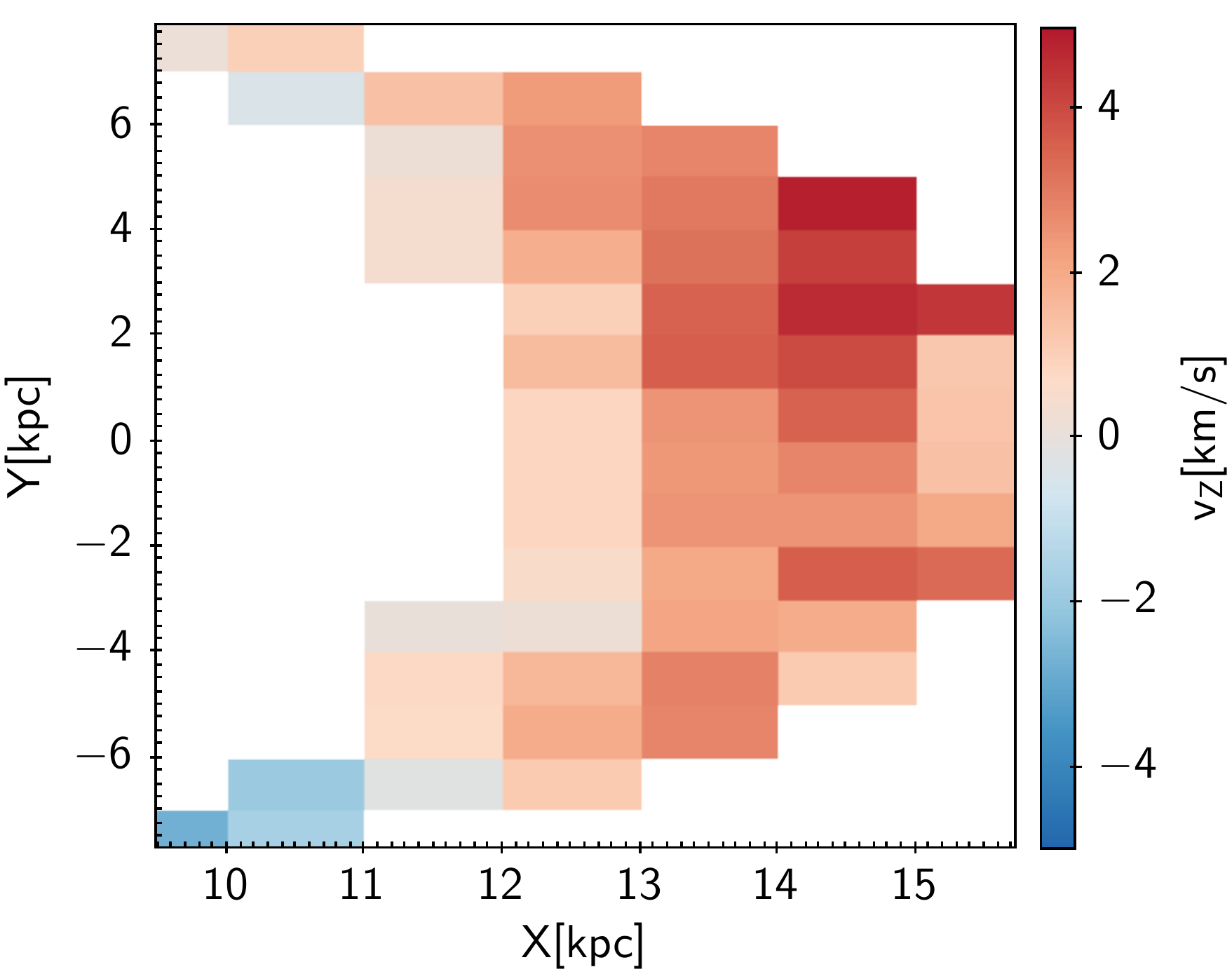}{0.35\textwidth}{(a)}
		\fig{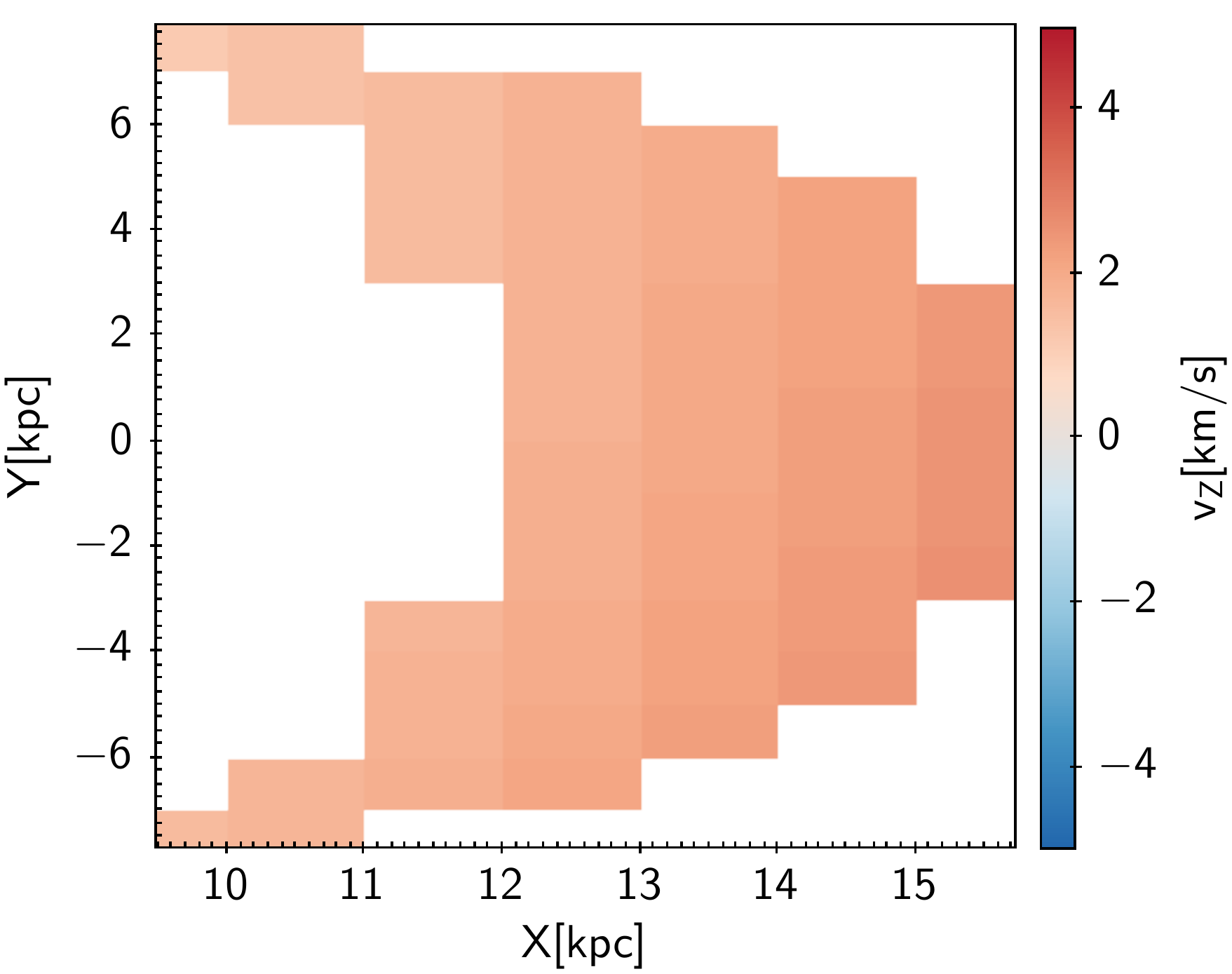}{0.35\textwidth}{(b)}
		\fig{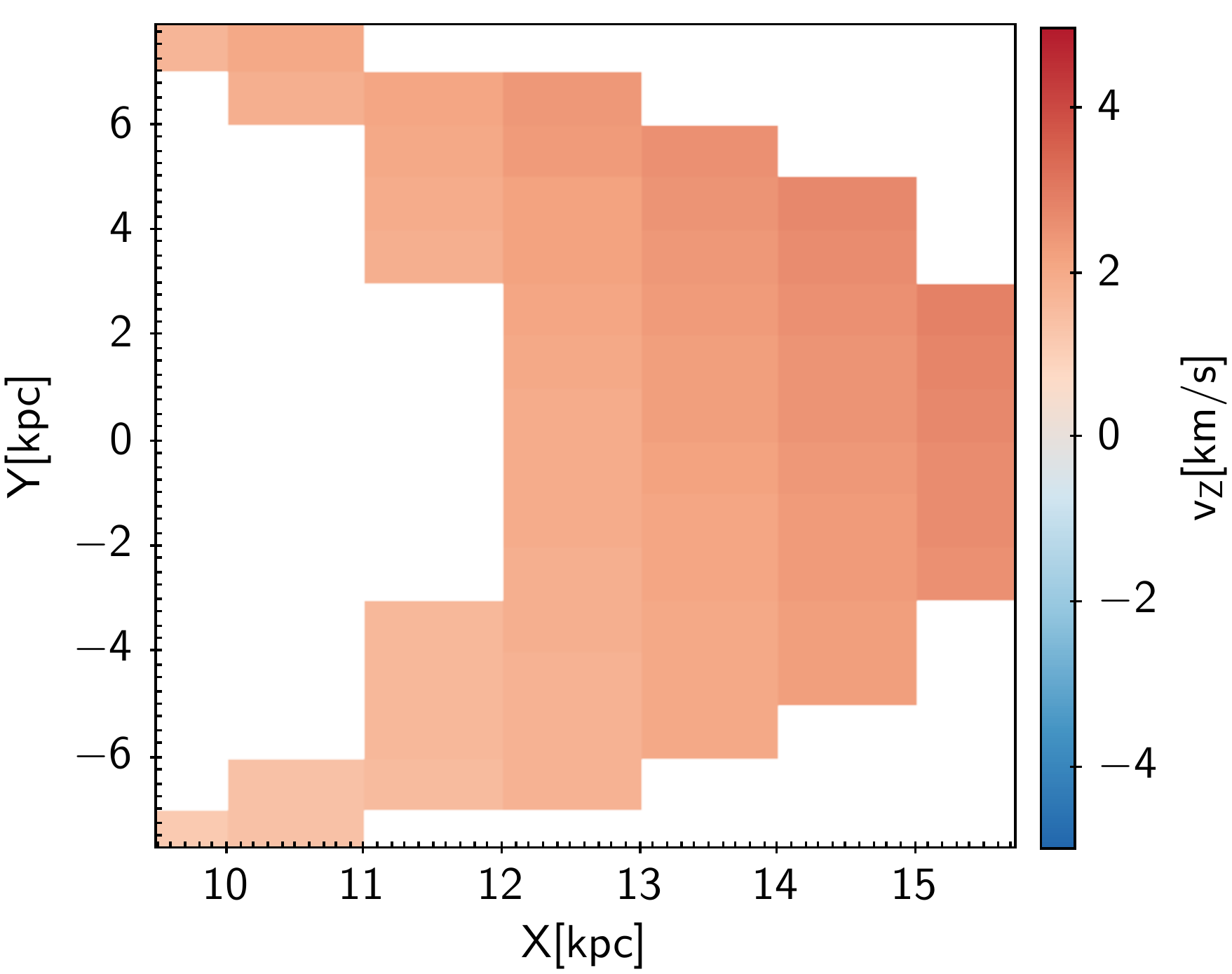}{0.35\textwidth}{(c)}
	}

	\gridline{\fig{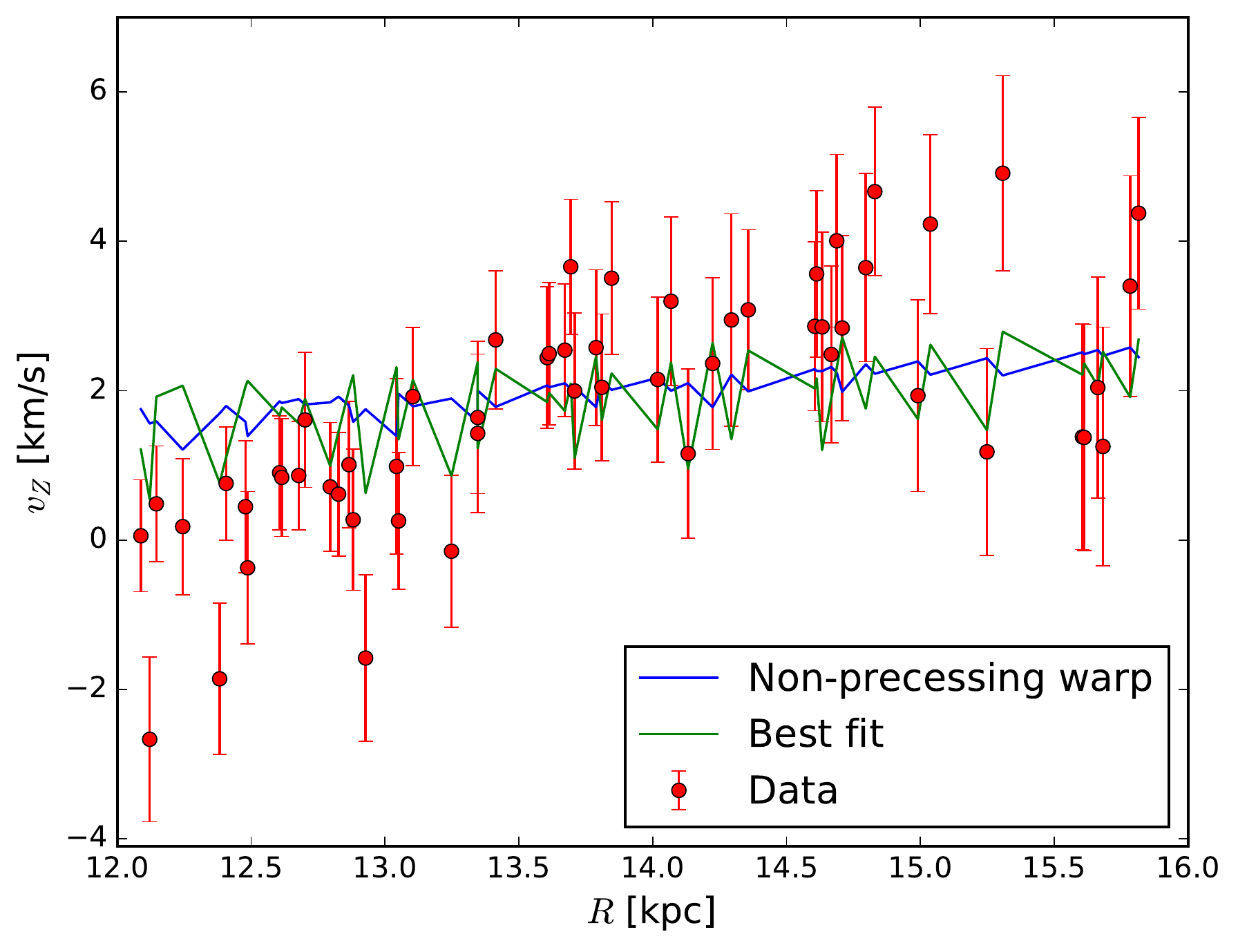}{0.35\textwidth}{(d)}
	\hspace{-2cm}
	\fig{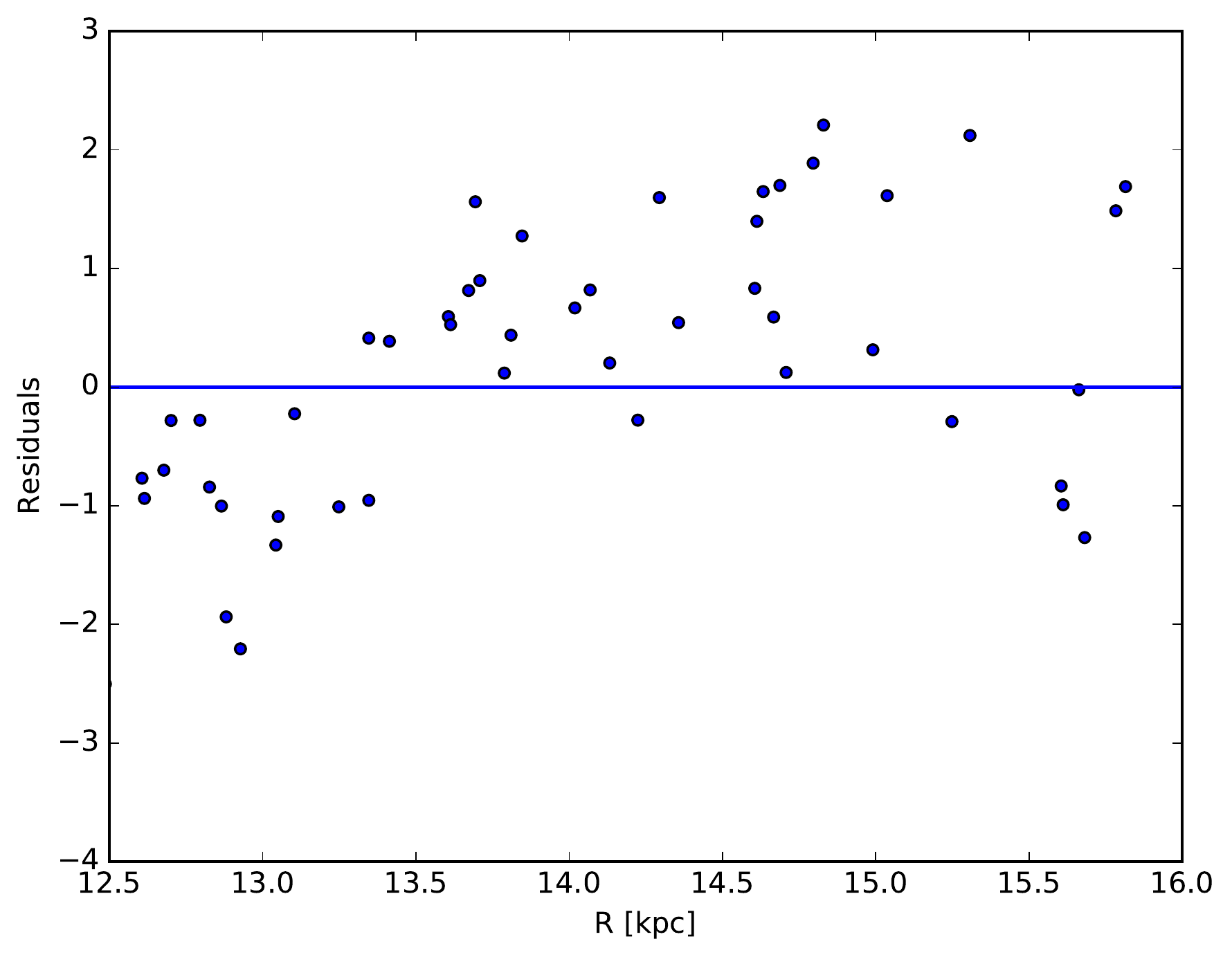}{0.35\textwidth}{(e)}
	}
	
	\caption{Result of fit using Equation (\ref{model}) with parameters of \cite{zofi}, fitting all data. (a) Data. (b) Nonprecessing steady warp ($\beta=0$ km s$^{-1}$ kpc$^{-1}$ and $K=0$ km s$^{-1}$ kpc$^{-1}$). (c) Best fit with $K$ and $\beta$ as free parameters. For the best fit, we find $K=16^{+12}_{-8}$ km s$^{-1}$ kpc$^{-1}$ and $\beta=4^{+6}_{-4}$ km s$^{-1}$ kpc$^{-1}$. (d) Two-dimensional representation of the fit. The blue curve is for the nonprecessing steady model of the warp, the green curve is the best fit of the data. (e) Residuals corresponding to the best fit.}
		\label{fig_final_zofi}
\end{figure*}

Next, we do the fit for all the azimuths, with the complete model (Equation (\ref{model})), including precession and time variation of the warp amplitude. Variation of the warp amplitude may in fact be more relevant than the precession term \citep{martin_ampl,martin_new}. For warp parameters of \cite{zofi}, we find the best fit to be for $K=16^{+12}_{-8}$ km s$^{-1}$ kpc$^{-1}$ and $\beta=4^{+6}_{-4}$ km s$^{-1}$ kpc$^{-1}$. For comparison, we did the same fit using the warp parameters of the linear model of \cite{chen} that \citetalias{poggio} used. This yields best-fit parameters $K=1.0^{+1.4}_{-2.2}$ km s$^{-1}$ kpc$^{-1}$ and $\beta=14.0 \pm 1.4$ km s$^{-1}$ kpc$^{-1}$. Similar numbers are obtained with other warp models used by \citetalias{poggio} or by calculating errors in a different way (see Table 1). The results of this fit can be seen in Figure \ref{fig_final_chen} and Figure \ref{fig_final_zofi}, where we find the same trend as in the anticenter. With the warp parameters of \cite{chen} there is the need for a high precession to fit the data, whereas when we use the warp model of \cite{zofi}, the nonprecessing steady warp is of the order of magnitude of the data. For the data, there is some substructure in velocity, but as we mention in the Section \ref{intro}, we are not attempting here to explain all the velocity substructures, but to analyze the overall effect of the warp. 

The difference between the warp model of \cite{chen} and the model of \cite{zofi} is that the first one was derived for a very young stellar population of Classical Cepheids, whereas the second one was calculated for the whole population traced by Gaia DR2, which is much older on average. Our results using the model of young population warp used by \citetalias{poggio} with our dataset are in agreement with their findings. However, when we apply warp parameters of the whole population, we cannot detect the precession. Since we have a large value of precession error, we cannot refute the result of \citetalias{poggio}, but we cannot reject the nonprecessing warp either. Since the error of the precession comes mostly from the propagation of the warp parameters (see Section \ref{errors}), we believe that the effect of error propagation is stronger than the error of the fits of velocities and we would need more precise data to be able to detect precession.

\subsection{Different Warp Models}
\citetalias{poggio} also use other warp parameters such as from \cite{yusifov} and \cite{martin_warp}. However, these models are not representative of the whole Gaia DR2 population either. The warp parameters of \cite{yusifov} were derived for pulsars only; therefore, they should not be applied to the whole population. Plus their parameters were derived only for Galactocentric distances $R\approx15$ kpc. Although \citetalias{poggio} argue that \cite{momany} found that the \cite{yusifov} model, describes the red-giant branch stars, a population similar to that of \citetalias{poggio} well, we do not think that that makes it applicable in the case of \citetalias{poggio} for the following reason. \cite{momany} do not have a 3D distribution of data, but a 2D projection at a fixed distance. Their data get well fitted by the \cite{yusifov} model because \cite{yusifov} measures a high value of the warp line of nodes, which describes the data projection of \cite{momany} well, but that does not imply that it describes the real 3D distribution of data as well. \cite{zofi} applied the \cite{yusifov} warp parameters on the whole Gaia DR2 dataset and we found that it leads to high warp amplitude, which is not in agreement with the warp amplitude of the whole population. 

The parameters of \cite{martin_warp} were derived for the red-clump stars, but these parameters were derived only for Galactocentric distances up to $R\approx 13$ kpc; therefore, extrapolating them at higher distances is unreliable and leads to unrealistic warp amplitude. 

In Figure \ref{modely_porovnanie}, we show that the maximum amplitude of models used by \citetalias{poggio} is quite similar, whereas the warp amplitude derived by  \cite{zofi} is lower by a factor of $\approx$ 3\textendash4. The result of fits with other models can be seen in Table \ref{tabulka}.

\subsection{Calculation of Error Bars}\label{errors}
Determining the error of $v_Z$ is not an easy task, as in the data, the systematic and statistical error are entangled and we cannot differentiate between them. Therefore, we tried several approaches to calculate the error. In each case, we normalize the error bars in order to obtain $\chi_r^2=1.0$ for the case of nonprecessing warp in the line of nodes, using warp parameters of \cite{zofi}. The factor $f$ in Table \ref{tabulka} is precisely introduced for this normalization, and it should be approximately $f=1/\sqrt{N}$ for the case of statistical errors and $f=1$ for the case of purely systematical errors. From the error of velocity, we can then calculate the error of fit parameters. 

\begin{figure*}
	\vspace{-0.3cm}
	\centering

	\includegraphics[width=0.4\textwidth]{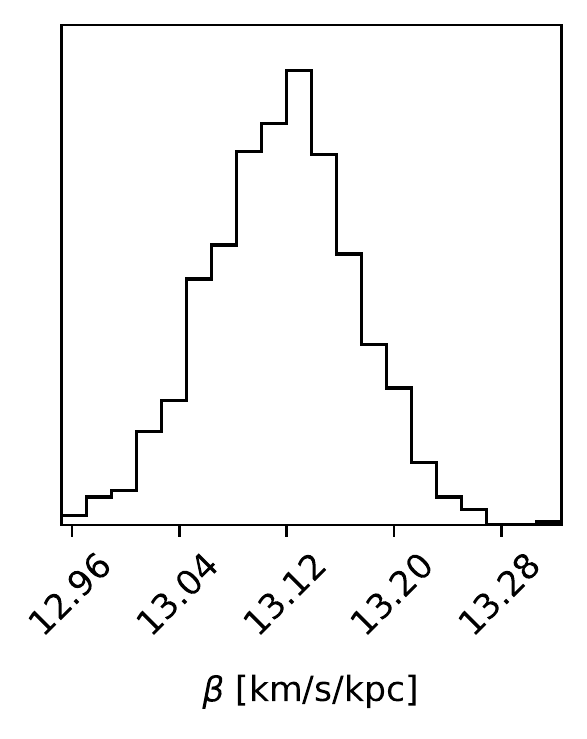}
	\includegraphics[width=0.385\textwidth]{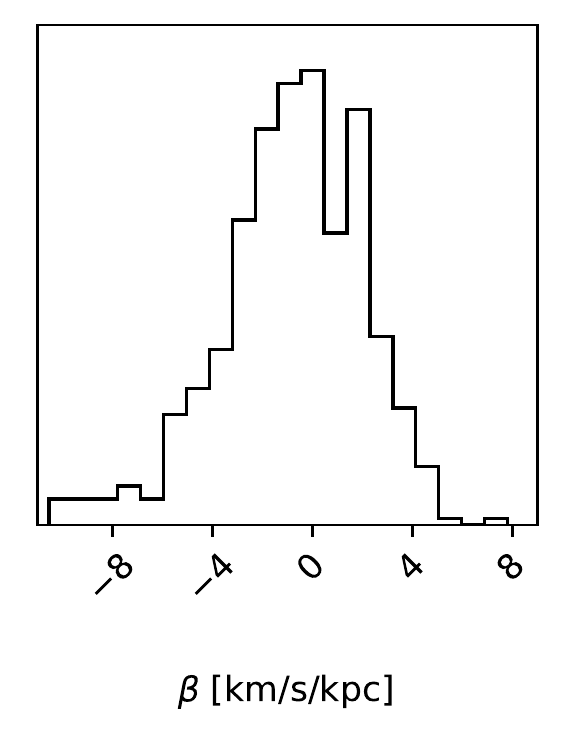}
	
	\caption{Corner plot of the Monte Carlo simulation using only data near the anticenter. Left: model of \citetalias{poggio}, with warp parameters of \cite{chen}. Right: model of \cite{zofi}.}\label{monte_carlo}
\end{figure*}

Another contribution to the error comes from the propagation of error of warp parameters. To determinate the error bars of $\beta$ and $K$ we made Monte Carlo simulations. The systematic and statistical error of the warp parameters (Equation (\ref{parametre})) was summed linearly. It is important to notice that for the warp model of \cite{zofi} this contribution to the error is the dominant one and causes the error bars to be significant. For illustration purposes, we show corner plots of Monte Carlo simulations for the linear model of \cite{chen} and \cite{zofi} in Figure \ref{monte_carlo}, using only data near the anticenter. We use the maximum spread of the values of $\beta$ to determine the error bars of the precession. It is clear that for the model of \cite{zofi} the spread of values of precession is very large, which impedes determining the precession more accurately, whereas for the model of \cite{chen}, the spread is far smaller.

Finally we summed quadratically the error of the fit and error produced by the uncertainties of warp parameters to determine the final error of the fit parameters $\beta$ and $K$. The values are given in Table \ref{tabulka}, where each model is given for a different determination of velocity error bar. These various approaches give similar results.

\begin{longrotatetable}
\begin{deluxetable}{p{.18\textwidth}p{.18\textwidth}p{.22\textwidth}p{.2\textwidth}p{.2\textwidth}p{.22\textwidth}}
	\tablenum{1}
	\tablecaption{Results of Fit with Various Warp Models and Error Calculations}
	\tablewidth{10pt}
	\tablehead{Model of Error & Conditions of the Fit & \cite{zofi} (km s$^{-1}$ kpc$^{-1}$)& \cite{chen}, Linear Model (km s$^{-1}$ kpc$^{-1}$) & \cite{chen}, Power Model (km s$^{-1}$ kpc$^{-1}$) & \cite{yusifov} (km s$^{-1}$ kpc$^{-1}$)}
		
	\startdata
	& $\ang{-10}<\phi<\ang{10}, \beta=0$ & $\chi^2_r=1.03$ & $\chi^2_r=81.54$ & $\chi^2_r=78.57$ & $\chi^2_r=62.51$ \\ 
	& $\ang{-10}<\phi<\ang{10}, \beta$ free & $\beta=-1 \pm 9, \chi^2_r=1.00$ & $\beta=13 \pm 1, \chi^2_r=1.8$ & $\beta=13 \pm 1, \chi^2_r=1.77$ & $\beta=13 \pm 1, \chi^2_r=1.83$  \\ 
	Model 1 & $\forall \phi, K=0, \beta=0$ & $\chi^2_r=1.94$ & $\chi^2_r=76.33$ & $\chi^2_r=73.55$  & $\chi^2_r=61.45$ \\
	& $\forall \phi,K=0, \beta$ free & $\beta=2^{+5}_{-8}, \chi^2_r=1.86$ & $\beta=14\pm 1, \chi^2_r=2.82$ & $\beta=14\pm 1, \chi^2_r=2.86$ & $\beta=14 \pm 1, \chi^2_r=2.88$   \\ 
	& $\forall \phi, \beta=0, K$ free & $K=11^{+17}_{-12}, \chi^2_r=1.75$ & $K=22 \pm 1, \chi^2_r=43.79$ & $K=22 \pm 1, \chi^2_r=41.98$ & $K=21 \pm 1, \chi^2_r=40.87$ \\ 
	& $\forall \phi, K$ free, $\beta$ free & $\beta=4^{+6}_{-4}, K=16^{+12}_{-8}, \chi^2_r=1.56$ & $\beta=14 \pm 1, K=1.0^{+1.4}_{-2.2}, \chi^2_r=2.83$ & $\beta=13 \pm 1, K=2.0 \pm 1.4, \chi^2_r=2.81$  & $\beta=13 \pm 1, K=2 \pm 1, \chi^2_r=2.85$ \\ 
	\cline{0-5}
	& $\ang{-10}<\phi<\ang{10}, \beta=0$ & $\chi^2_r=1.05$ & $\chi^2_r=84.70$ & $\chi^2_r=81.79$ & $\chi^2_r=64.74$ \\ 
	& $\ang{-10}<\phi<\ang{10}, \beta$ free & $\beta=-1^{+7}_{-9}, \chi^2_r=1.00$ & $\beta=13 \pm 1, \chi^2_r=1.71$ & $\beta=13 \pm 1, \chi^2_r=1.68$ & $\beta=13 \pm 1, \chi^2_r=1.71$  \\ 
	Model 2 & $\forall \phi, K=0, \beta=0$ & $\chi^2_r=2.30$ & $\chi^2_r=90.32$ & $\chi^2_r=87.23$ & $\chi^2_r=73.74$ \\ 
	& $\forall \phi,K=0, \beta$ free & $\beta=3^{+5}_{-8}, \chi^2_r=2.09$ & $\beta=14\pm 1, \chi^2_r=3.31$ & $\beta=14\pm 1, \chi^2_r=3.34$ & $\beta=14 \pm 1, \chi^2_r=3.40$ \\ 
	& $\forall \phi, \beta=0, K$ free & $K=8^{+14}_{-8}, \chi^2_r=2.18$ & $K=21 \pm 1, \chi^2_r=54.77$ & $K=21 \pm 1, \chi^2_r=52.62$ & $K=20 \pm 1, \chi^2_r=51.19$ \\ 
	& $\forall \phi, K$ free, $\beta$ free & $\beta=6^{+4}_{-6}, K=15^{+11}_{-7}, \chi^2_r=1.79$ & $\beta=14 \pm 1, K=1.0 \pm 1.4, \chi^2_r=3.23$ & $\beta=14 \pm 1, K=1.0 \pm 1.4, \chi^2_r=3.21$ & $\beta=14 \pm 1, K=2^{+1}_{-2}, \chi^2_r=3.26$  \\ 
	\cline{0-5}
	& $\ang{-10}<\phi<\ang{10}, \beta=0$ & $\chi^2_r=1.93$ & $\chi^2_r=81.83$ & $\chi^2_r=78.99$ & $\chi^2_r=61.57$ \\ 
	& $\ang{-10}<\phi<\ang{10}, \beta$ free & $\beta=-2^{+10}_{-8}, \chi^2_r=1.00$ & $\beta=13 \pm 1, \chi^2_r=1.54$ & $\beta=13 \pm 1, \chi^2_r=1.53$ & $\beta=13 \pm 1, \chi^2_r=1.63$ \\ 
	Model 3 & $\forall \phi, K=0, \beta=0$ & $\chi^2_r=2.05$ & $\chi^2_r=77.50$ & $\chi^2_r=74.69$ & $\chi^2_r=61.72$ \\ 
	& $\forall \phi,K=0, \beta$ free & $\beta=1^{+7}_{-10}, \chi^2_r=1.99$ & $\beta=14\pm 1, \chi^2_r=2.91$ & $\beta=14 \pm 1, \chi^2_r=2.96$ & $\beta=14 \pm 1, \chi^2_r=3.01$ \\ 
	& $\forall \phi, \beta=0, K$ free & $K=12 \pm 13, \chi^2_r=1.80$ & $K=21 \pm 1, \chi^2_r=45.22$ & $K=21 \pm 1, \chi^2_r=43.34$ & $K=21 \pm 1, \chi^2_r=41.65$ \\ 
	& $\forall \phi, K$ free, $\beta$ free & $\beta=4^{+5}_{-4}, K=17^{+7}_{-8}, \chi^2_r=1.63$ & $\beta=13 \pm 1, K=2.0 \pm 1.4, \chi^2_r=2.83$ & $\beta=13 \pm 1, K=2.0 \pm 1.4, \chi^2_r=2.80$ & $\beta=13 \pm 1, K=2 \pm 1, \chi^2_r=2.85$ \\ 
	\enddata
	\tablecomments{	Model 1: error calculated as $\sigma^2_i=f\cdot \frac{\sum_{j} w_j \sigma_{Z,j}^2}{\sum_{j} w_j}$, quadratically combined with the error propagated from warp parameters. Model 2: error calculated as $\sigma^2_i=f\cdot \frac{\sum_{j} w_j (v_{Z,j}-\overline{v_{Z,i}})^2}{\sum_{j} w_j}$, quadratically combined with the error propagated from warp parameters. Model 3: error calculated as $\sigma^2_i=f\cdot \frac{1}{\sqrt{\sum_{j} \frac{1}{\sigma_{Z,j}^2}}}\sqrt{\frac{\chi_b^2}{N-1}}, \chi_b^2=\sum_{j}\frac{(v_{Z,j}-\overline{v_{Z,i}})^2}{\sigma_{Z,j}^2}$, quadratically combined with the error propagated from warp parameters. For the warp model of \cite{zofi}, the propagation of warp parameters is the dominant component of the error. In all cases, $f$ is such that $\chi^2_r$[$\ang{-10}<\phi<\ang{10}$; \cite{zofi}]$=1.00$ for the best fit.}
	\label{tabulka}
\end{deluxetable}
\end{longrotatetable}

\subsection{Non-Gaussian Residuals of Least-squares Fit}
\begin{figure}
	\begin{center}
		\includegraphics[width=0.5\textwidth]{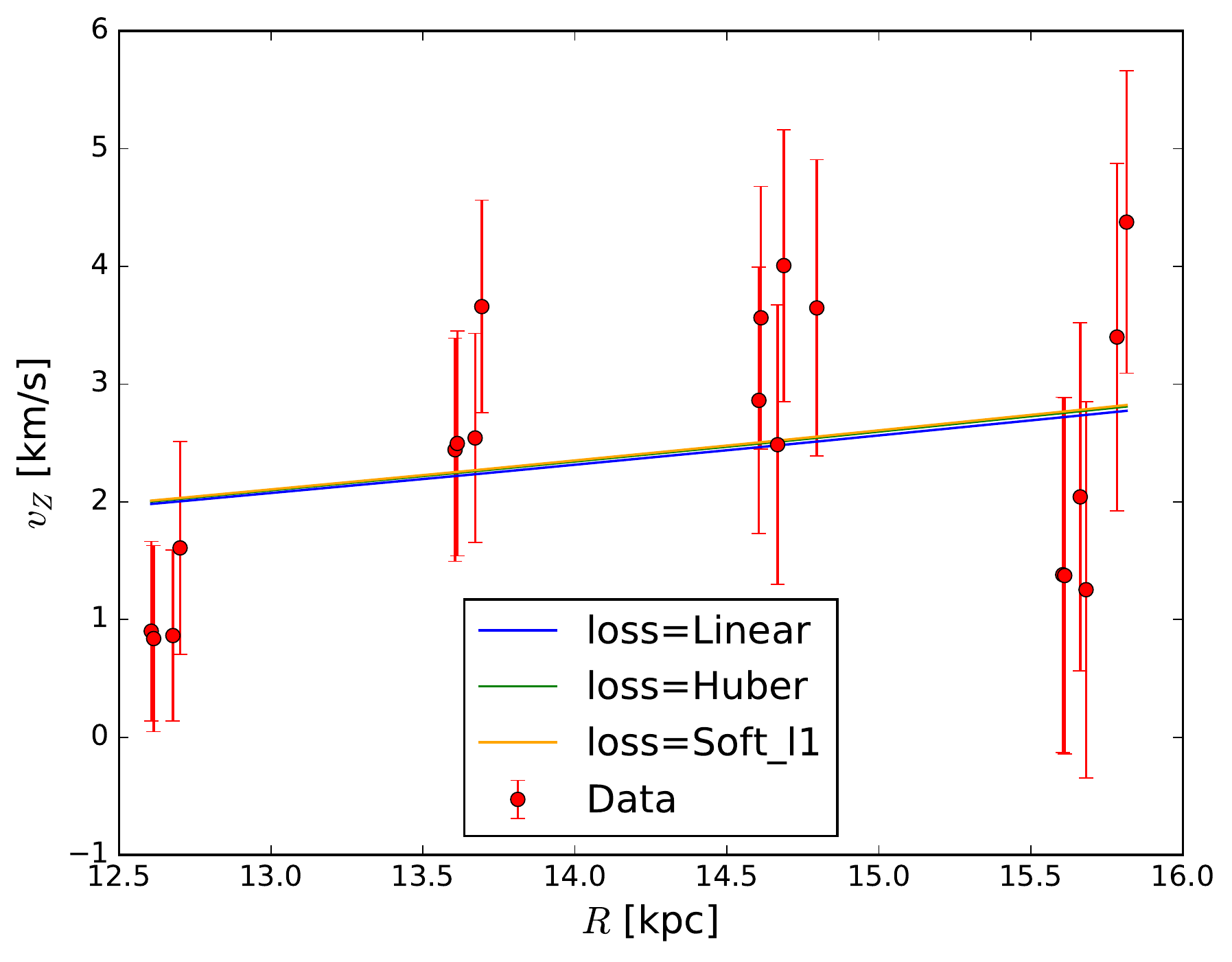}
		\caption{Comparison of the least-squares fit using different loss functions to treat outliers. We fit only data near the anticenter and we apply the warp model of \cite{zofi}. The linear loss function corresponds to the ordinary least-squares fit, where we obtain the value of precession $\beta=-1\pm1.8$ km s$^{-1}$ kpc$^{-1}$. With a smooth approximation of l1 we obtain $\beta=-1.35\pm2.21$ km s$^{-1}$ kpc$^{-1}$ and with Huber's function we obtain $\beta=-1.07\pm1.81$ km s$^{-1}$ kpc$^{-1}$. Note that in this case we only report the error of precession coming from the fit, not combined with the propagation of error of warp parameters, which is why the error is lower than in previous sections.}\label{fit_losses}
	\end{center}
\end{figure}

In Figures \ref{lon}, \ref{fig_final_chen}, and \ref{fig_final_zofi}, we can see that the residuals of the least-squares fitting are not completely Gaussian, especially for the case of the warp model of \cite{zofi} the residuals are skewed. To account for this effect, we performed one of the fits using Python's method curve fit from the Scipy package. This fitting method has implemented several loss functions that we can apply to perform a more robust fit with less sensitivity to outliers. In Figure \ref{fit_losses} we show the comparison of the fit using a linear loss function that corresponds to the standard least-squares fit with more robust loss functions such as Huber's loss function and a smooth approximation of l1, which uses the sum of absolute deviations instead of squared ones (see Scipy documentation for more details). Change in the loss function influences the value of precession and the corresponding error, but compared to the error from the propagation of the warp parameters, this difference is insignificant; therefore, in all the cases we apply traditional least-squares as described in Section \ref{methods}.

\subsection{Comparison with \cite{cheng}}
\begin{figure}
	\begin{center}
		\includegraphics[width=0.5\textwidth]{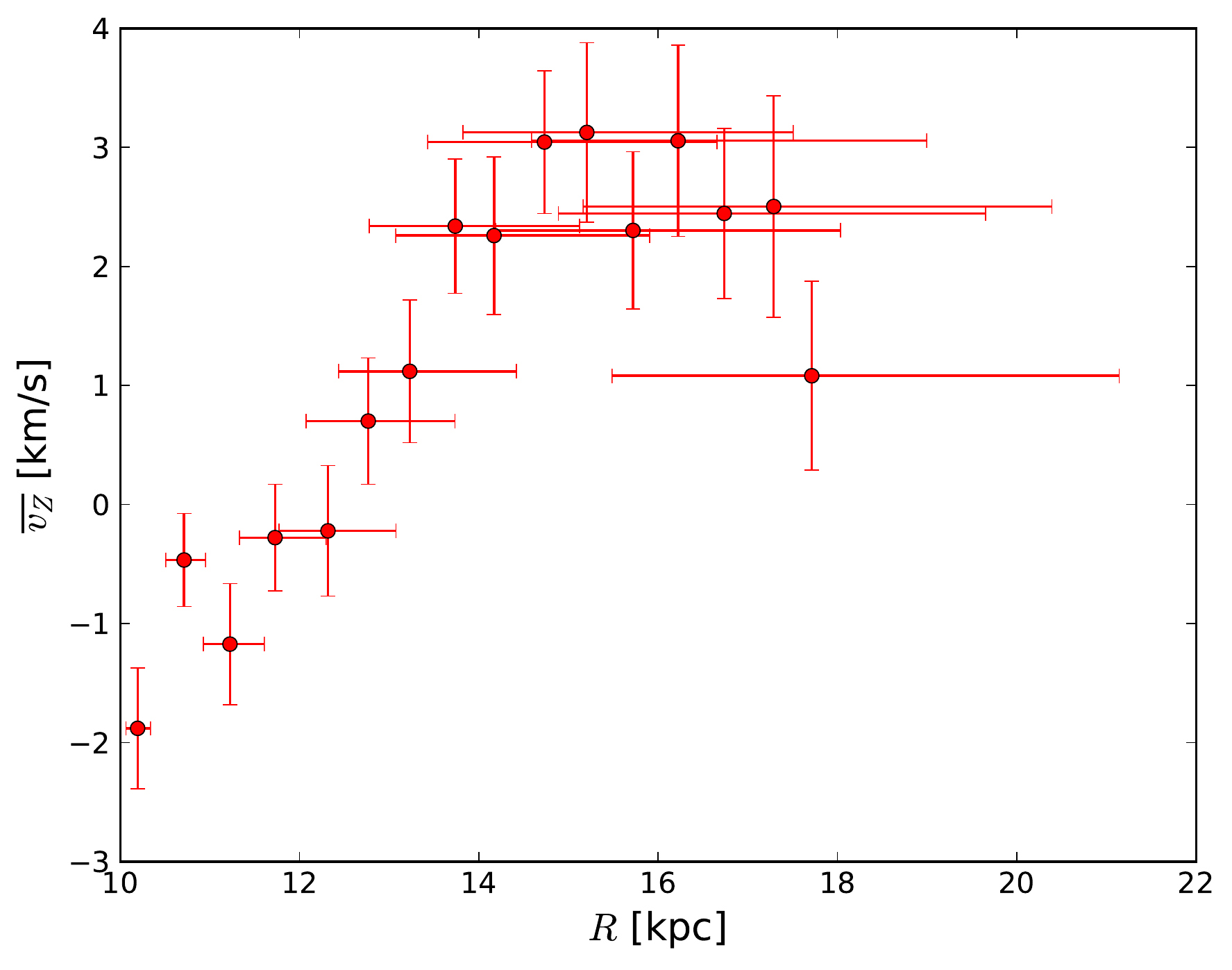}
		\caption{Average vertical velocity from Gaia DR2 including the complete range of azimuths. We do not observe a significant drop in velocity until $R \approx 17$ kpc. Due to large error bars, we constrain our analysis to $R<16$ kpc}\label{vz_priemer}
	\end{center}
\end{figure}

The claims of \citetalias{poggio} are supported in a recent paper of \citet[hereafter C20]{cheng}, who used Gaia DR2 and APOGEE-2 DR16 to fit the warp from the kinematics. They use distances derived through Bayesian inference using the StarHorse code \citep{queiroz}, which enables them to reach a high Galactocentric distance of $R=18$ kpc. They observe a significant decrease in vertical velocity, which they attribute to the warp. They adopt a model including the precession of warp and leave all the warp parameters free, except the line of nodes. They find a high amplitude of warp and a high value of warp precession of $\beta=13.57^{+0.20}_{-0.18}$ km s$^{-1}$ kpc$^{-1}$, similar to the value of \citetalias{poggio}. This is in disagreement with our results, due to the drop in the vertical velocity, beginning at about $R \approx 13$ kpc. We do not observe this in our data. As can be seen in Figure \ref{vz_priemer}, our data show a decrease in vertical velocity, starting at about $R \approx 17$ kpc. We are aware that our statistical method to calculate the distance can give rise to errors; however, from the horizontal error bars in Figure \ref{vz_priemer} we can see that up to $R \approx 16$ kpc the data is robust and the error of the distance determination starts to be significant at higher distances. Therefore even taking into account the error bars, the drop in velocity would manifest itself at higher distances than given by \citetalias{cheng}.

\begin{figure}
	\begin{center}
		\includegraphics[width=0.5\textwidth]{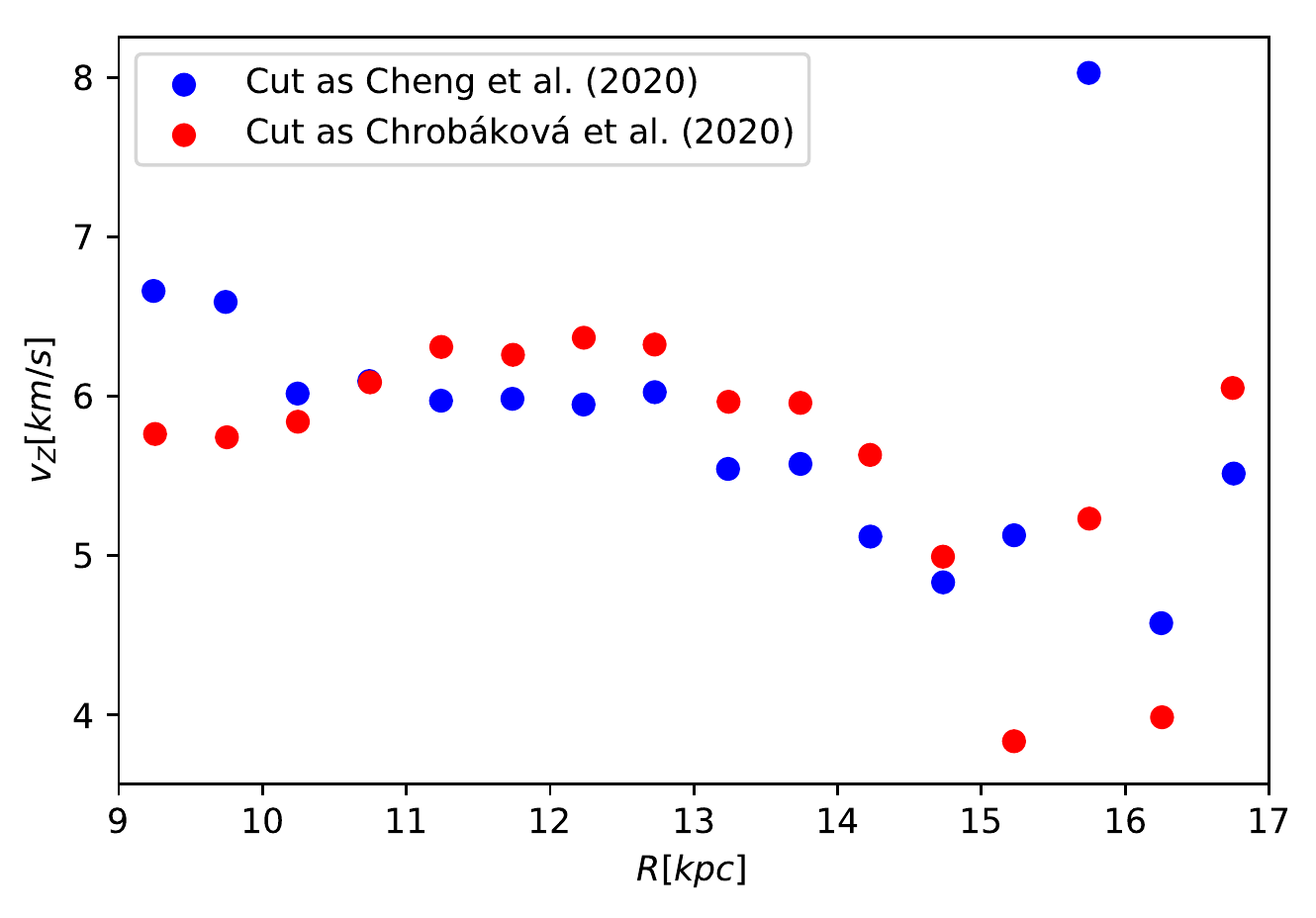}
		\caption{Average vertical velocity from Gaia DR2. For the blue curve we applied the same constraints as \citetalias{cheng}, most importantly a cut in [Fe/H] and [Mg/Fe] to choose only thin disk stars. For the red curve, we applied the same constraint as in our data, that is $\lvert b \rvert < \ang{10}$. Contrary to \citetalias{cheng}, we do not observe a significant drop in velocity for either approach.}\label{vz_starhorse}
	\end{center}
\end{figure}

As a further attempt to find the origin of the discrepancy in the data, we repeated the analysis carried out by \citetalias{cheng} with similar data. \citetalias{cheng} use two different datasets. One is a combination of Gaia DR2 with StarHorse distances. We are not aware that \citetalias{cheng} put any constraints on these data, except filtering out stars that have a problematic Gaia photometric or astrometric solution or a troublesome StarHorse data reduction and removing stars from the Large Magellanic Cloud (LMC) and Small Magellanic Cloud (SMC). Therefore we have to assume they include all ranges of Galactic latitudes in this sample, thus not avoiding contamination by thick disk and halo stars. The second dataset of \citetalias{cheng} combines Gaia DR2 with chemical and radial velocity information from APOGEE-2 DR16 and with StarHorse distances. \citetalias{cheng} apply a constraint to this dataset in [Fe/H] and [Mg/Fe] in order to choose only thin disk stars (see their Figure 1). For both datasets, \citetalias{cheng} observe a drop in velocities, although for the second sample the drop is less prominent. The first sample of \citetalias{cheng} should be similar to what we are using, except we only chose stars with $\lvert b \rvert < \ang{10}$, which would explain the discrepancy. To reproduce the second sample, we used a publicly available dataset of \cite{queiroz_data} who combined spectroscopic data from APOGEE-2 DR16 with photometric data and parallaxes from Gaia DR2 and distances derived using Starhorse. We applied the same selection criteria as \citetalias{cheng} \textemdash we removed sources within $\ang{5}$ from the center of LMC and SMC, we restricted the effective temperature between $3700$ K and $5500$ K, and we applied the same constraint in [Fe/H] and [Mg/Fe]. Our dataset is not exactly the same as the one of \citetalias{cheng}, but they should be in good agreement. However, upon applying the same conditions as \citetalias{cheng}, we do not observe a drop in velocities as \citetalias{cheng} obtained. In fact, the velocity that we obtain is consistent with the velocity obtained when we applied our condition on the data, that is $\lvert b \rvert < \ang{10}$ (see Figure \ref{vz_starhorse}). 
Therefore we do not know what the reason is behind the drop in vertical velocity found by \citetalias{cheng}, but we suspect there might be contamination by the thick disk and the halo stars. In order to fit the warp, \citetalias{cheng} used the Gaia sample that was not constrained and \citetalias{cheng} state they assume this sample is not contaminated because the velocity has similar behavior to that of the thin disk in the Gaia-APOGEE sample. Although in principle this method of constraining the thin disk should be correct, we doubt that a full Gaia DR2 sample chosen within $\lvert b \rvert = \ang{90}$ has the same behavior of velocity as a strictly constrained thin disk sample of APOGEE. The fact that we cannot reproduce the drop in velocity at $R\approx13$ kpc in either sample confirms this doubt, although we do not know with certainty what the reason is for this discrepancy. 

\citetalias{cheng} explain the drop in velocities solely in terms of warp and use only the vertical velocity to fit the warp parameters. We plot their warp amplitude in Figure \ref{modely_porovnanie}, where we can see that their model yields far higher values than other works. We do not know if this result would be sustained by their stellar density, as \citetalias{cheng} did not analyze the density distribution. However, \cite{zofi} calculated warp parameters by fitting the stellar density and reaching a significantly lower warp amplitude.

Our argument is supported by other results in the literature as well, since most authors obtain far lower warp amplitude, between roughly $0.4-1$ kpc at $R=15$ kpc, depending on the stellar population and warp model \citep{yusifov,reyle,chen,li,skowron}. The only work that supports the result of \cite{cheng} is the one of \cite{amores}, who produced star counts using a population synthesis model (Besançon Galaxy Model); therefore, we think their results should be interpreted with care since they are model-\text{dependent}.

\section{Conclusions}
We studied the precession of the Galactic warp based on the model of \citetalias{poggio} and present our own calculation. We are using vertical velocities of Gaia DR2 that represent the average kinematics of an old stellar population ($\sim$5-6 Gyr) and we apply a new warp model \citep{zofi}, based on Gaia DR2. Our warp model has a significantly smaller amplitude than models used by \citetalias{poggio}, which are either derived for a young stellar population ($\sim$Myr) or an invalid extrapolation of an old stellar population warp. When we consider variation of the amplitude of the warp with the age of the population, we obtain a best fit that is compatible with no precession, taking into account the uncertainties in a warp model independently derived from an old stellar density distribution with the same Gaia data. Therefore we do not find evidence to support or exclude any model of warp formation. Future studies are necessary to understand kinematics of warp and thus its formation mechanism. The next Gaia data release DR3 could provide sufficiently precise data, to study the warp in greater detail and possibly detect the precession.

\acknowledgments
We thank E. Puha, H.-F. Wang, and R. Nagy for helpful comments. \v{Z}.C. and M.L.C. were supported by the grant PGC-2018-102249-B-100 of the Spanish Ministry of Economy and Competitiveness (MINECO). This work has made use of data from the European Space Agency (ESA) mission Gaia (\url{https://www.cosmos.esa.int/gaia}), processed by the Gaia Data Processing and Analysis Consortium (DPAC, \url{https://www.cosmos.esa.int/web/gaia/dpac/consortium}). Funding for the DPAC has been provided by national institutions, in particular the institutions participating in the Gaia Multilateral Agreement. Funding for the Sloan Digital Sky 
Survey IV has been provided by the 
Alfred P. Sloan Foundation, the U.S. 
Department of Energy Office of 
Science, and the Participating 
Institutions. 

SDSS-IV acknowledges support and 
resources from the Center for High 
Performance Computing  at the 
University of Utah. The SDSS 
website is www.sdss.org.

SDSS-IV is managed by the 
Astrophysical Research Consortium 
for the Participating Institutions 
of the SDSS Collaboration including 
the Brazilian Participation Group, 
the Carnegie Institution for Science, 
Carnegie Mellon University, Center for 
Astrophysics | Harvard \& 
Smithsonian, the Chilean Participation 
Group, the French Participation Group, 
Instituto de Astrof\'isica de 
Canarias, The Johns Hopkins 
University, Kavli Institute for the 
Physics and Mathematics of the 
Universe (IPMU) / University of 
Tokyo, the Korean Participation Group, 
Lawrence Berkeley National Laboratory, 
Leibniz Institut f\"ur Astrophysik 
Potsdam (AIP),  Max-Planck-Institut 
f\"ur Astronomie (MPIA Heidelberg), 
Max-Planck-Institut f\"ur 
Astrophysik (MPA Garching), 
Max-Planck-Institut f\"ur 
Extraterrestrische Physik (MPE), 
National Astronomical Observatories of 
China, New Mexico State University, 
New York University, University of 
Notre Dame, Observat\'ario 
Nacional / MCTI, The Ohio State 
University, Pennsylvania State 
University, Shanghai 
Astronomical Observatory, United 
Kingdom Participation Group, 
Universidad Nacional Aut\'onoma 
de M\'exico, University of Arizona, 
University of Colorado Boulder, 
University of Oxford, University of 
Portsmouth, University of Utah, 
University of Virginia, University 
of Washington, University of 
Wisconsin, Vanderbilt University, 
and Yale University.

\bibliographystyle{aa} 
\bibliography{Precesia}
	
\end{document}